\documentclass[final,5p,times]{elsarticle}
\usepackage{seqsplit}
\usepackage[T1]{fontenc}
\usepackage[utf8]{inputenc}
\usepackage{amsmath,amsfonts}
\usepackage{algorithmic}
\usepackage{algorithm}
\usepackage{array}
\usepackage{booktabs}
\usepackage{textcomp}
\usepackage{stfloats}
\usepackage[hyphens]{url}
\usepackage{verbatim}
\usepackage{graphicx}
\usepackage{multirow}
\usepackage{amssymb}
\usepackage{stfloats}
\usepackage{diagbox}
\usepackage{amsmath,amsfonts}
\usepackage[table,xcdraw]{xcolor}
\usepackage{amsthm}
\usepackage{subcaption} 
\usepackage[normalem]{ulem}
\useunder{\uline}{\ul}{}
\newtheorem{definition}{Definition}[section]

\usepackage{graphicx}

\usepackage{listings}
\usepackage[most]{tcolorbox}
\usepackage{enumitem}
\usepackage{hyperref}
\hypersetup{
    colorlinks=true,
    allcolors=blue,
    pdftitle={Software monitors for control-flow anomaly detection through large language models and conformance checking},
    pdfpagemode=FullScreen,
}

\def\nomove{\gg}

\usepackage[table]{xcolor} 
\usepackage{listings}
\usepackage[T1]{fontenc}

\tcbset{
  myNiceBox/.style={
    colback=gray!5,
    colframe=gray!60!black,
    coltext=black,
    rounded corners,
    boxrule=0.8pt,
    left=1em,
    right=1em,
    top=0.5em,
    bottom=0.5em,
    width=\linewidth,
    breakable,
    before skip=0.5\baselineskip,
    after skip=0.5\baselineskip,
  }
}
\newtcolorbox{prompt}[1][]{
  myNiceBox,
  title=\textbf{#1}
}
\newtcolorbox{excerpt}[1][]{
  myNiceBox,
  title=\textbf{#1}
}
\newtcolorbox{rquestion}[1][]{
  myNiceBox,
  title=\textbf{#1}
}

\definecolor{mybackcolor}{rgb}{0.98, 0.98, 0.98}
\lstdefinestyle{pythonstyle}{
    language=Python,
    basicstyle=\ttfamily\small\linespread{0.9}\selectfont,
    keywordstyle=\color{blue},
    stringstyle=\color{red!60!black},
    commentstyle=\color{green!40!black},
    morekeywords={self, True, False, None, return, if, elif, else, not, and, or, in, def, class, assert},
    keywordstyle=[2]{\color{purple}},
    emph={DataStatus, Level, Mode, RadioNetwork, ERTMSOnBoardSystem, MockController},
    emphstyle=[1]{\color{teal}\bfseries},
    emph={[2] _phase_1_get_initial_data, _procedure_s1_driver_id_entry, _procedure_d2_check_pos_level,
          _procedure_d3_check_level_valid, _procedure_s2_level_entry, _simulate_driver_action,
          test_path_l2_success_grant_fs, setup_test_logger, run_start_of_mission},
    emphstyle={[2]\color{brown}},
    showstringspaces=false,
    frame=single,
    framerule=0.2pt,
    rulesepcolor=\color{gray!50},
    captionpos=b,
    breaklines=true,           
    breakatwhitespace=false,   
    breakautoindent=true,
    breakindent=0pt,
    columns=flexible,          
    postbreak=\mbox{\textcolor{red}{$\hookrightarrow$}\space},
    backgroundcolor=\color{mybackcolor},
    tabsize=2,
    rulecolor=\color{black!30},
    framesep=10pt, 
    xleftmargin=5pt,
    xrightmargin=5pt,
    aboveskip=5pt,
    belowskip=5pt
}

\journal{Elsevier}

\begin{document}
\begin{frontmatter}

\title{Architecting software monitors for control-flow anomaly detection through\\ large language models and conformance checking}

\author[federicoii]{Francesco Vitale}
\author[supsi,unifi]{Francesco Flammini}
\author[linnaeus]{Mauro Caporuscio}
\author[federicoii]{Nicola Mazzocca}
\affiliation[federicoii]{organization={University of Naples Federico II}, addressline={Via Claudio, 21}, city={Naples}, postcode={80125}, country={Italy}}
\affiliation[supsi]{organization={University of Applied Sciences and Arts of Southern Switzerland}, addressline={Via la Santa 1}, city={Lugano}, postcode={6962}, country={Switzerland}}
\affiliation[unifi]{organization={University of Florence}, addressline={Viale Morgagni, 67}, city={Florence}, postcode={50134}, country={Italy}}
\affiliation[linnaeus]{organization={Linnaeus University}, addressline={Universitetsplatsen, 1}, city={V\"{a}xj\"{o}}, postcode={35252}, country={Sweden}}

\begin{abstract}
\textbf{Context:} Ensuring high levels of dependability in modern computer-based systems has become increasingly challenging due to their complexity. Although systems are validated at design time, their behavior can be different at runtime, possibly showing control-flow anomalies due to ``unknown unknowns''.\\
\textbf{Objective:} We aim to detect control-flow anomalies through software monitoring, which verifies runtime behavior by logging software execution and detecting deviations from expected control flow. \\
\textbf{Methods:} We propose a methodology to develop software monitors for control-flow anomaly detection through Large Language Models (LLMs) and conformance checking. The methodology builds on existing software development practices to maintain traditional V\&V while providing an additional level of robustness and trustworthiness. It leverages LLMs to link design-time models and implementation code, automating source-code instrumentation. The resulting event logs are analyzed via conformance checking, an explainable and effective technique for control-flow anomaly detection.\\
\textbf{Results:} We test the methodology on a case-study scenario from the European Railway Traffic Management System / European Train Control System (ERTMS/ETCS), which is a railway standard for modern interoperable railways. The results obtained from the ERTMS/ETCS case study demonstrate that LLM-based source-code instrumentation can achieve up to 82.849\% control-flow coverage of the reference design-time process model, while the subsequent conformance checking-based anomaly detection reaches a peak performance of 95.957\% F1-score and 93.669\% AUC.\\
\textbf{Conclusion:} Incorporating domain-specific knowledge to guide LLMs in source-code instrumentation significantly allowed obtaining reliable and quality software logs and enabled effective control-flow anomaly detection through conformance checking.

\end{abstract}

\end{frontmatter}

\section{Introduction}
\label{sec:intro}
The complexity of modern computer-based systems increases their exposure to several types of threats, including software defects, hardware faults, and malicious attacks \cite{huang2016softwareresilience, karaduman2023cpsresilience}. 
The usage of model-based verification and formal methods can significantly help in fault avoidance and detection \cite{vigano2023ddcpsmutationtesting, bombarda2022ratemodelbasedtesting, casaluce2024statisticalmodelchecking}; however, practical limitations in test coverage and scalability of formal methods, as well as changes and uncertainties in the system and its environments, make verification and validation extremely challenging.
These limitations pose risks to computer-based systems' dependability \cite{pradhan2016dssresilience, pan2023softwareresilienceempiricalstudy}. 
We aim to demonstrate that a higher level of dependability can be achieved by software monitoring, which involves logging the software execution and detecting any deviations from the expected behavior \cite{cinque2016characterizingdirectmonitoring, gu2023softwareloggingpractices}. Specifically, we focus on \textit{control-flow anomaly detection}, which deals with the identification of misalignments between the expected flow of activities and their actual sequencing \cite{vitale2025cfad}.

Software monitoring has been implemented through runtime verification \cite{cinque2013rulebasedmonitoring, kejstova2017modelcheckingruntimeverification, cimatti2019nurv, czepa2020temporalpropertiesunderstandability}.
Runtime verification is closely aligned with formal software specifications and can accurately identify errors that violate them, although its effectiveness depends on the completeness and correctness of the underlying models. In fact, runtime verification requires the synthesis of software monitors by specifying properties and assumptions about the execution, involving a thorough understanding of the software functionality, strong observability guarantees, and control over the environment in which the software is deployed~\cite{cimatti2025effectivemonitoring}. These challenges are particularly pronounced in modern software systems, where maintaining strict consistency between system models and their implementations is difficult due to system complexity and multi-developer settings \cite{jongeling2022consistencymanagement,weidmann2018tolerant,trols2019multifaceted}. Moreover, increasingly complex systems introduce incompleteness and uncertainty, including so-called ``unknown unknowns''~\cite{sinclair2025chasing} and, more specifically, residual faults~\cite{natella2013sfi}, which cause anomalous behavior that cannot be anticipated due to the lack of prior knowledge of their existence. Such emergent unwanted behavior can have multiple causes, such as system integration of different, possibly commercial-off-the-shelf components or lack of resources to evaluate all the possible normal and faulty scenarios of the system \cite{tetlie2023systemsintegration, prokop2024aerospacesoftwareerrors}. 

To account for unforeseen behaviors and environmental uncertainty at runtime, control-flow anomaly detection can exploit machine/deep learning approaches to characterize the observable normal behavior directly from log data and use such characterization to check for deviations at runtime~\cite{singh2017fuzzyfaultprediction, du2017deeplog, denaro2023softwarefailuresmonitoring, han2023loggpt, guan2024wake, zhang2025llmbasedadexplaination}. While these approaches are more flexible and domain-agnostic, they suffer from significant and broadly documented limitations related to their explainability and trustworthiness, which hinder their deployment in critical scenarios~\cite{rawal2022trustworthyaiadvances}. Moreover, their data-driven and domain-agnostic nature loosens the relationship with the software specification, which makes connecting their diagnoses with high-level designs more challenging.

\begin{figure*}[!t]
\centering
\includegraphics[width=\textwidth]{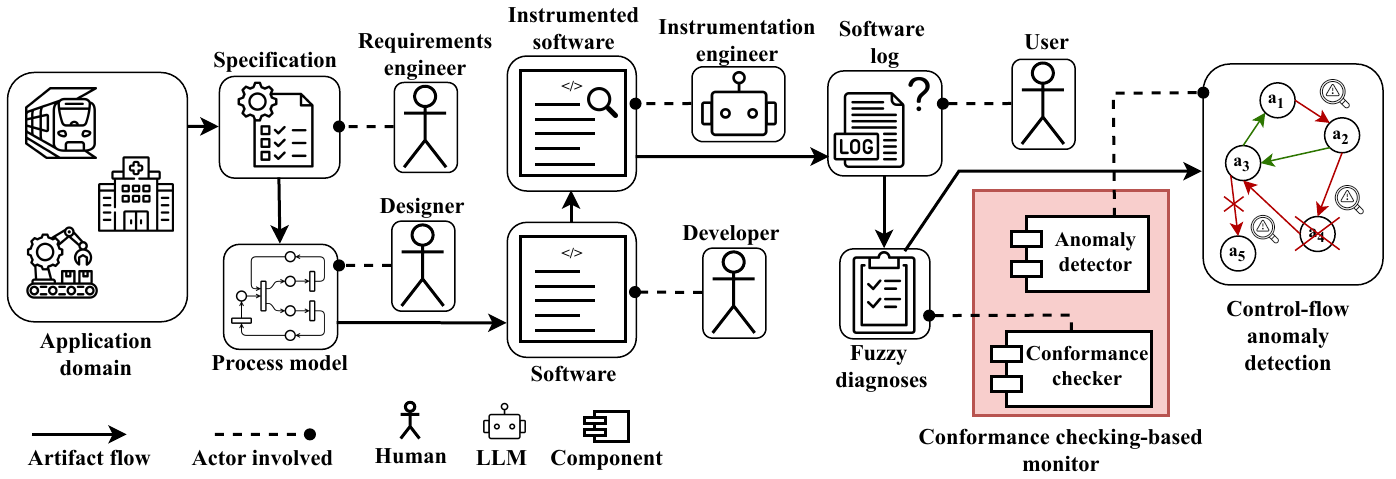}
\caption{High-level view of the methodology and the conformance checking-based monitor. Humans are involved in the generation of the specification of the application domain, the description of the process model capturing the software behavior, and the implementation of the software. The instrumented software is generated by an LLM by bridging the high-level design with the low-level implementation. Users interact with the instrumented software, which generates logs throughout its execution. These logs are automatically analyzed by a conformance checker, which provides fuzzy diagnoses. Finally, an anomaly detector uses the fuzzy diagnoses to verify the presence of control-flow anomalies.}
\label{PM_FLOW}
\end{figure*}

In this paper, we address 1) the limitations of traditional runtime verification in terms of its difficulties in capturing all possible properties and assumptions about the monitored system, and 2) the trustworthiness of machine/deep learning control-flow anomaly detection approaches and their lack of connection with high-level design. In particular, our proposal combines a new instrumentation technique capable of automatically connecting software implementations with high-level designs and conformance checking, an effective and explainable method for control-flow anomaly detection. The instrumentation technique is based on Large Language Models (LLMs). LLMs are advanced neural networks trained on vast amounts of code and textual data, enabling them to understand software structure and control flow \cite{havare2025codecomprehensionbenchmarklarge}. By leveraging these capabilities, LLMs can automatically generate instrumentation code and allow obtaining high-quality software logs, which is a key requirement for quality conformance checking results~\cite{martin2021pmopportunitieschallenges, zimmermann2023pmanalysischallenges, mamudu2024pmsuccessfactors}.

Combining LLMs and conformance checking, we formulate the following two Research Questions (RQs):

\begin{itemize}
    \item \textbf{RQ1 (Source-code instrumentation)}: How can LLMs be leveraged to bridge high-level design models and software implementations and generate high-quality logs for subsequent monitoring through conformance checking?
    \item \textbf{RQ2 (Control-flow anomaly detection)}: How effective is conformance checking-based control-flow anomaly detection in identifying software anomalies using LLM-enabled software instrumentation?
\end{itemize}

To address these RQs, we propose a methodology enabling the development of conformance checking-based monitors, whose high-level view is shown in Figure \ref{PM_FLOW}. Based on the application domain, a requirements engineer develops a specification. This is used to drive the design of a prescriptive process model and the development of the software. The software is instrumented by an LLM, which bridges the high-level design with the low-level structure and control-flow of the software. The software log is generated when a user interacts with the system and compared with the process model through a conformance checker. This results in fuzzy diagnoses that an anomaly detector processes to find any control-flow anomalies. In conclusion, our proposal addresses the challenge of bridging software implementations with high-level software design through LLM-based software instrumentation, and allows for the flexible alignment and deviation identification using these designs through conformance checking.

We test the methodology and the conformance checking-based monitor on a case-study scenario from the European Rail Traffic Management System / European Train Control System (ERTMS/ETCS), a railway standard setting the specification for software development and testing of modern interoperable railways \cite{laroche2013ertms}. The ERTMS/ETCS requirements specification document, SUBSET-026, prescribes the behavior of train components in all reference operational scenarios\footnote{\url{https://www.era.europa.eu/era-folder/1-ccs-tsi-appendix-mandatory-specifications-etcs-b4-r1-rmr-gsm-r-b1-mr1-frmcs-b0-ato-b1}}. The results obtained from the ERTMS/ETCS case study demonstrate that LLM-based source-code instrumentation can achieve up to 84.755\% control-flow coverage, while the subsequent conformance checking–based anomaly detection reaches a peak performance of 96.610\% F1-score and 93.515\% AUC.

In summary, the novel contributions of this paper are:
\begin{itemize}
    \item A fuzzy and explainable conformance checking–based runtime monitor capable of detecting control-flow anomalies from software logs generated by LLM-instrumented code.
    \item A methodology to guide LLM-enabled source-code instrumentation and development of the fuzzy and explainable conformance checking-based monitor for software monitoring of computer-based systems.
    \item The application of the methodology and the evaluation of the monitor capabilities referencing the ERTMS/ETCS standard, a real-world case study documenting the specification of European railways.
\end{itemize}

The rest of the paper is organized as follows. Section \ref{sec:motivation} motivates the research work through a detailed description of ERTMS/ETCS systems. Section \ref{sec:related_work} presents the related works on software logging and anomaly detection.
Section \ref{sec:cc_based_monitor} describes the conformance checking-based monitor architecture. Section \ref{sec:methodology} describes the methodology driving the engineering and use of conformance checking-based software monitoring.
Section \ref{sec:case_study} describes the case-study scenario from the ERTMS/ETCS standard, the techniques employed for control-flow anomaly detection, and the experimental factors and metrics. Section \ref{sec:evaluation} presents the results. Section \ref{sec:discussion} discusses the results as responses to the research questions, and presents the threats to validity. Section \ref{sec:conclusions} concludes and hints about future research directions. 
\section{Motivation}
\label{sec:motivation}
Modern computer systems will run within increasingly variable and unpredictable environments. A running example are railway systems, which are characterized by many sources of uncertainty, such as environmental conditions, railway traffic, and connection issues between on-board and trackside equipment \cite{ERA2024SafetyReport}. In this paper, we target the safety-critical ERTMS/ETCS railway systems, which control the operation of trans-European railway traffic. 

\begin{figure}[!t]
\centering
\includegraphics[width=\columnwidth]{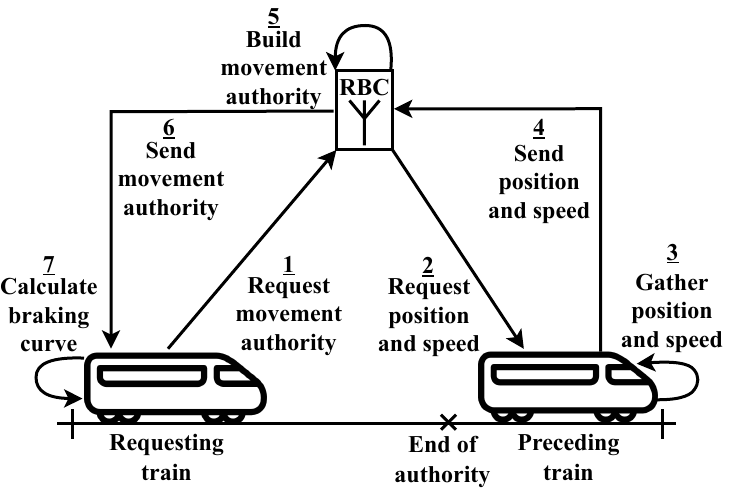}
\caption{Request of the movement authority and calculation of the braking curve by a train in ERTMS/ETCS systems.}
\label{ERTMS_MA_PROVISION}
\end{figure}
An ERTMS/ETCS railway system involves the digital elaboration of data generated by trackside and on-board equipment \cite{terbeek2018ertmsetcscps}. Essential on-board components are the Driver Machine Interface (DMI), European Vital Computer (EVC), Balise Transmission Module (BTM), and the Radio Transmission Module (RTM): the DMI enables the driver to engage with ERTMS/ETCS procedures and control the train; the EVC implements the needed logic for safely managing data flow within the on-board subsystem and between on-track and on-board communications; the BTM reads information through Eurobalises; and the RTM handles all the needed train-to-infrastructure communication logic. On the trackside, the RBC supervises trains in its area and handles several functionalities through the interlocking system to ensure efficient and safe train operation. The type of processing depends on the level of operation, from level 1 to level 3, which defines the degree of automation and the needed trackside and on-board equipment. 
At level 1, the system uses active Eurobalises to transmit signals to the train, providing discontinuous signalling. Level 2 introduces continuous radio communication between the train and trackside equipment, enabling more efficient operation.
Level 3 adds moving block signalling and train integrity check equipment for even higher capacity.
Let us consider an example ERTMS/ETCS critical task: the acquisition of the movement authority by a train. The movement authority provides information about the distance from other trains to the EVC of a train. The EVC must calculate the braking curve based on such information and on the static speed profiles to ensure the safety of the passengers and infrastructure. Figure \ref{ERTMS_MA_PROVISION} shows the acquisition of the movement authority in ERTMS/ETCS level 2 and level 3, which enables the calculation of the braking curve of the requesting train by letting the RBC collect the train positions wirelessly.

The case-study ERTMS/ETCS scenario that we target in our experiments is the Start of Mission (SoM) procedure. This procedure is rather complex and critical, as its goal regards obtaining a movement authority at the start of a journey. The procedure includes: validating the driver's identity; registering to the closest RBC for train supervision; reporting the train position; obtaining railway traffic information; and obtaining the movement authority. 

Although SoM is described in the ERTMS/ETCS system requirements specification, it is mostly specified in natural language and can be subject to interpretation ambiguity among different manufacturers of system components. In fact, extensive field integration testing involving different companies is usually required, which however cannot exclude interaction anomalies in unspecified situations and edge cases. Software monitoring can identify these anomalies by tracing and checking key information regarding software executions. In the next section, we are exploring related works on software monitoring and outlining the key novelty of this work's proposal in comparison to existing approaches.

\begin{table*}[!t]
\centering
\caption{Comparison of the reviewed literature in terms of instrumentation of software logging and anomaly detection. The symbol ``---'' indicates that the reference did not detail or include the related functionality. ML = Machine Learning, RV = Runtime Verification, PM = Process Mining.}
\label{tab:RELATED_COMPARISON}
\resizebox{\textwidth}{!}{%
\begin{tabular}{l|lll|lll}
\hline
\textbf{Ref.}                                         & \multicolumn{3}{l|}{\textbf{Logging}}                                                  & \multicolumn{3}{l}{\textbf{Anomaly detection}}                                 \\ \cline{2-7} 
\textbf{}                                             & \textbf{Type} & \textbf{Flexible} & \textbf{Log semantics}                             & \textbf{Type}    & \textbf{Fuzzy} & \textbf{Explainable}                       \\ \hline
\cite{briand2006umlsequencediagramreverseengineering} & Source code   & No                & High-level (object interactions)                   & ---              & ---            & ---                                        \\
\cite{cinque2013rulebasedmonitoring}                  & Source code   & No                & High-level (service interactions)                  & RV               & No             & Yes (intrinsic)                            \\
\cite{zhang2015javacodeinstrumentation}               & Source code   & No                & High-level (object interactions)                   & ---              & ---            & ---                                        \\
\cite{kejstova2017modelcheckingruntimeverification}   & Indirect      & Yes               & Low-level (system calls)                           & RV               & No             & Yes (intrinsic)                            \\
\cite{singh2017fuzzyfaultprediction}                  & ---           & ---               & ---                                                & ML               & Yes            & Yes (intrinsic)                            \\
\cite{du2017deeplog}                                  & ---           & ---               & ---                                                & ML               & Yes            & Partly (post-processing)                   \\
\cite{jia2018smartlog}                                & ML            & Yes               & High-level (explicit messages)                     & ---              & ---            & ---                                        \\
\cite{cinque2019hiddenapplicationerrorspm}            & ---           & ---               & ---                                                & PM               & Yes            & Yes (intrinsic)                            \\
\cite{cimatti2019nurv}                                & Source code   & No                & Low-level (software variables)                     & RV               & No             & Yes (intrinsic)                            \\
\cite{pecchia2020applicationfailuresanalysispm}       & ---           & ---               & ---                                                & PM               & Yes            & Yes (intrinsic)                            \\
\cite{engelke2020instrew}                             & Binary code   & Yes               & Low-level (system calls, memory access)            & ---              & ---            & ---                                        \\
\cite{denaro2023softwarefailuresmonitoring}           & ---           & ---               & ---                                                & ML               & Yes            & No                                         \\
\cite{han2023loggpt}                                  & ---           & ---               & ---                                                & ML               & Yes            & No                                         \\
\cite{guan2024wake}                                   & ---           & ---               & ---                                                & ML               & Yes            & Partly (post-processing)                   \\
\cite{mastropaolo2024logstatementdeeplearning}        & ML            & Yes               & High-level (explicit messages)                     & ---              & ---            & ---                                        \\
\cite{wang2025autolog}                                & ML            & Yes               & High-level (explicit messages)                     & ---              & ---            & ---                                        \\
\cite{zhang2025llmbasedadexplaination}                & ---           & ---               & ---                                                & ML               & Yes            & Partly (post-processing)                   \\
\cite{vitale2025cfad}                                 & ---           & ---               & ---                                                & PM + ML          & Yes            & Yes (intrinsic + post-processing)          \\
\cite{FlexInstru2026}                                 & Binary code   & Yes               & Low-level (branches, instructions)                 & ---              & ---            & ---                                        \\ \hline
\textbf{This work}                                    & \textbf{ML}   & \textbf{Yes}      & \textbf{High-level (design-time model activities)} & \textbf{PM + ML} & \textbf{Yes}   & \textbf{Yes (intrinsic + post-processing)} \\ \hline
\end{tabular}%
}
\end{table*}
\section{Related work}
\label{sec:related_work}
This section provides an overview of the existing work regarding software monitoring, which we split into software logging and software anomaly detection.
\subsection{Software logging}
\paragraph{Binary-code instrumentation}The literature has proposed many binary-code instrumentation frameworks. Engelke and Schulz \cite{engelke2020instrew} presented Instrew, which is similar to Valgrind, but optimizes instrumentation intrusiveness by leveraging the LLVM compiler infrastructure. Mu et al.~\cite{FlexInstru2026} developed FlexInstru, which introduces a hardware-independent dynamic instrumentation framework. By leveraging a novel process attachment/detachment mechanism and an instrumentation sampling strategy, FlexInstru allows for flexible control over instrumentation duration, effectively balancing profiling accuracy with runtime overhead.
While binary-code instrumentation is very flexible and can be optimized to reduce its overhead regarding software execution time and program size, source-code instrumentation typically leads to more interpretable and higher-quality logs. This is pivotal to obtaining meaningful descriptions of software executions. 

\paragraph{Source-code instrumentation}Briand et al. \cite{briand2006umlsequencediagramreverseengineering} proposed the instrumentation of Java programs to record objects' interactions and log the sequences of messages exchanged. Their objective was to reverse engineer the program and automatically obtain a sequence diagram to enhance the understanding of the software's internals. Cinque et al. \cite{cinque2013rulebasedmonitoring} introduced the rule-based source-code instrumentation technique, which involves placing logging instructions in specific places in the program according to a system representation developed at design time. This allows for the collection of high-quality logs linked to the entities of the system and their interactions. 
Zhang et al. \cite{zhang2015javacodeinstrumentation} proposed a declarative approach to tag the structures of Java programs and insert logging instructions corresponding to those structures. The approach requires querying the program and specifying instrumentation commands, which can include logging instructions. 

\paragraph{Machine learning-based instrumentation} Recent research has also leveraged machine learning to automate the placement and generation of log statements. Jia et al. \cite{jia2018smartlog} utilized random-forest classifiers to identify necessary logging points within exception blocks. Addressing content generation, Mastropaolo et al. \cite{mastropaolo2024logstatementdeeplearning} employed T5 transformers to automatically produce complete log statements based on the surrounding source code. Most recently, Wang et al. \cite{wang2025autolog} applied LLMs via in-context learning to both locate logging opportunities and generate descriptive messages, achieving higher accuracy than previous deep learning approaches.

\subsection{Software anomaly detection}
\paragraph{Runtime verification}Traditional approaches to identifying any anomalous behavior in software involve runtime verification of declarative specifications. Cinque et al. \cite{cinque2013rulebasedmonitoring} developed rule-based source-code instrumentation to verify whether the modeled software interactions between objects and internal services failed due to missing variable initialization, wrong value assigned to a variable, or missing function call. Cimatti et al.~\cite{cimatti2019nurv} presented an LTL-based runtime verification framework where a reference model is used to define assumptions and constraints on the software's behavior. Kejstová et al. \cite{kejstova2017modelcheckingruntimeverification} extended the idea of runtime verification by proposing runtime model-checking frameworks. The framework involves collecting the software state during its execution to restrict the state space of the reference software model. In this way, model-checking of declarative specifications becomes feasible at runtime due to the reduced time needed to check the specification properties. It is worth noting that the mentioned works may exploit some form of software instrumentation (e.g., source-code instrumentation) or monitor the software indirectly through, e.g., its system calls.

\paragraph{Machine learning-based anomaly detection}Although runtime verification is effective for verifying software behavior, it requires accurate software specifications and models. Hence, the literature put forward several data-driven solutions for software anomaly detection. For example, Singh et al. \cite{singh2017fuzzyfaultprediction} proposed a rule-based framework for anomaly detection that involves the selection of an optimized set of software features and the extraction of fuzzy rules that discriminate correct behaviors from anomalous ones. Denaro et al. \cite{denaro2023softwarefailuresmonitoring} proposed a neural network-based unsupervised anomaly detection framework. The solution employs deep autoencoders trained using software key performance indicators to discriminate whether the software execution is correct. It is worth noting that domain-agnostic data-driven methods grounded in deep learning with extension to the software domain have also been proposed. For example, Du et al.~\cite{du2017deeplog} have proposed DeepLog, a method based on long-short term memory networks, which are a type of recurrent neural network. The architecture is able to capture deviations between the predicted and observed events, and detect anomalies in system execution. Extensions to recurrent neural networks are those integrating the reconstruction-based anomaly detection paradigm through autoencoders, such as the method proposed in~\cite{guan2024wake}. Recently, LLMs have also been adopted for anomaly detection in system logs. Han et al.~\cite{han2023loggpt} introduced LogGPT, which is an LLM-based framework that involves LLM pre-training, reinforcement learning-based fine-tuning, and online inference to predict whether new log sequences are normal. Zhang et al.~\cite{zhang2025llmbasedadexplaination} extended this paradigm to include a human-in-the-loop explanation process.

\paragraph{Process mining-based anomaly detection}Although data-driven approaches have proven effective, they require laborious feature extraction through ad-hoc selection of software metrics and/or trace encodings. In addition, the limited trustworthiness and explainability of advanced machine learning limit its use. A promising area of research that addresses these issues is process mining, which combines traditional model-based analyses with data-driven insights. Cinque et al. \cite{cinque2019hiddenapplicationerrorspm} evaluated the capability of process mining to discover the normal behavior of the Apache web server as Petri nets with several algorithms, including the $\alpha$-miner, inductive miner, and integer linear programming-based miner. Subsequently, the authors performed a sensitivity analysis for the threshold to assign to the fitness metric for discriminating normal and anomalous behavior. Similarly, Pecchia et al. \cite{pecchia2020applicationfailuresanalysispm} evaluated the anomaly detection capability of process mining. The sensitivity analysis related to thresholding fitness assessed that higher fitness values lead to improved detection performance. However, the authors also remark that the quality of the model being discovered is critical to quality detection performance. In fact, they correlated noise in event logs handled by process discovery with negative effects on performance. In particular, as noise increases in event logs, there is a higher chance of false negatives, i.e., more misclassification of anomalies. Vitale et al. \cite{vitale2025cfad} developed a domain-agnostic process mining-based framework that performs feature extraction through conformance checking diagnoses and integrates an explainable machine learning component that exploits such diagnoses for control-flow anomaly detection.

\subsection{Our approach}
Table \ref{tab:RELATED_COMPARISON} compares our work to the reviewed literature with respect to both software logging and anomaly detection. From the comparison, the main novelty of our approach is enabling conformance checking-based control-flow anomaly detection through LLM-based instrumentation. In fact, by utilizing LLMs to align code logic with design-time model semantics, we automate the generation of logging lines that are not only syntactically correct but also semantically meaningful and highly interpretable. This allows extending the process mining-based control-flow anomaly detection framework proposed in~\cite{vitale2025cfad} to the software domain.

\begin{figure*}[!t]
\centering
\includegraphics[width=0.9\textwidth]{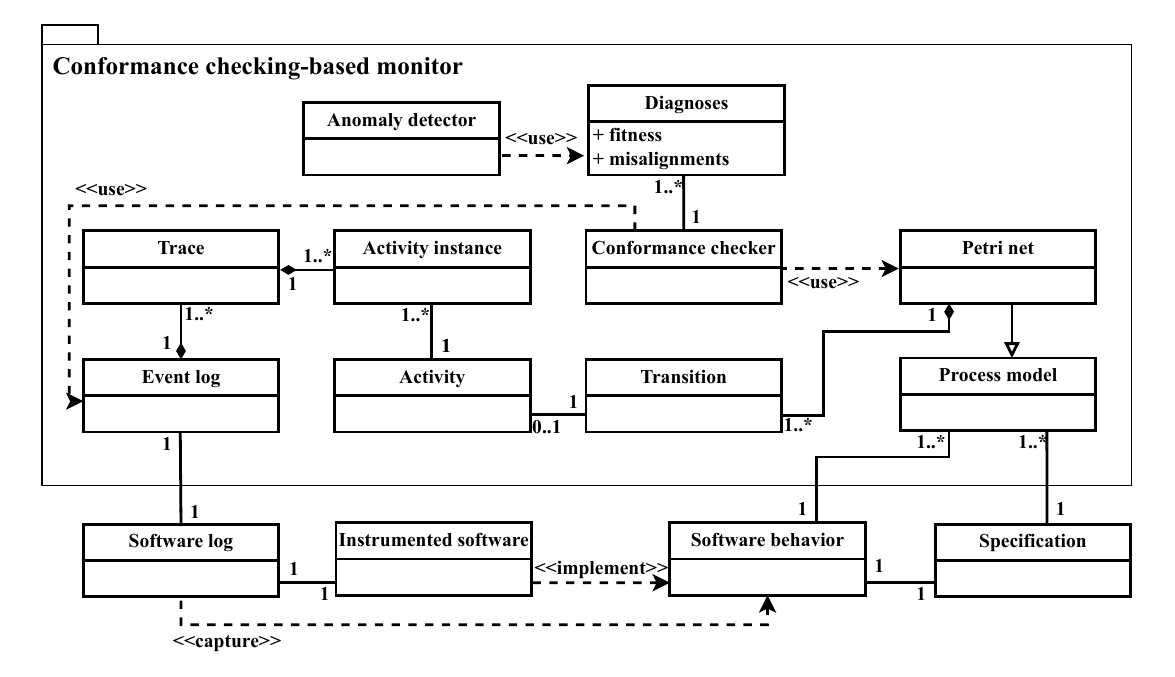}
\caption{Conceptual UML class diagram of the conformance checking-based monitor and other related methodological artifacts.}
\label{MONITOR_ARCH}
\end{figure*}
\section{The conformance checking-based monitor architecture}
\label{sec:cc_based_monitor}
The conformance checking-based monitor connects well-known concepts of the process mining community and software artifacts produced in traditional software development processes. On the one hand, process mining includes the concepts of process models and event logs, the two fundamental inputs to conformance checking. On the other hand, the software artifacts allow generating these inputs, hence their presence is pivotal to the application of conformance checking. Figure \ref{MONITOR_ARCH} shows a conceptual UML class diagram of the overall architecture. In this section, we are interested in describing the monitor architecture. The other concepts will be explored in detail in Section \ref{sec:methodology}.

The main goal of the monitor is to generate diagnoses through the \textbf{conformance checker} and use these diagnoses to detect control-flow anomalies through the \textbf{anomaly detector}. 

\subsection{Event log}
Conformance checking algorithms use event logs, which are structured data that collect uniquely identified activity instances of a reference process \cite{aalst2022pmhb}. In this paper, we focus on sequences of activity instances, i.e., traces; thus, we consider the following definition:
\begin{definition}[Event log]
\label{def:event_log}
Let $\mathcal{A}$ denote the universe of activities. Let $\mathcal{A}^*$ denote the universe of traces. An example trace is $\langle \sigma_1, \sigma_2,\dots,\sigma_n\rangle$. Let $\mathcal{B}(\mathcal{A}^*)$ indicate the set of multisets of traces. An event log is an element of $\mathcal{B}(\mathcal{A}^*)$, i.e., $L\in\mathcal{B}(\mathcal{A}^*)$. An example event log is 
\begin{flalign*}
L &= [\langle\sigma_1,\dots,\sigma_\alpha\rangle^a, \langle\sigma_1,\dots,\sigma_\beta\rangle^b, \langle\sigma_1,\dots,\sigma_\gamma\rangle^c],&
\end{flalign*}
where $\sigma_\alpha,\sigma_\beta,\sigma_\gamma$ indicate the last activity instance within the corresponding trace, and $a,b,c$ indicate the number of times the three traces appear in the event log.
\end{definition}

The event log is associated with a software log recorded from the execution of the software instrumented through LLMs. The criticality lies in maintaining a reliable association between the activities of the event log and the process model. This will be explored in more detail in the instrumentation phase of the methodology detailed in Section \ref{sec:methodology}.

\subsection{Process model}
Event logs are checked against a process model, which is a behavioral model that captures control-flow relationships between activities. The most common formalism employed by process mining is the Petri net, and, in particular, we focus on the labeled accepting Petri net variant \cite{aalst2022pmhb}.
\begin{definition}[Labeled accepting Petri net]
\label{def:petri_net}
Let $P$ and $Tr$ be two sets of nodes of a bipartite graph such that $P\,\cap\,Tr=\emptyset$, and $F\subseteq (P\times Tr)\,\cup\,(Tr\times P)$ a set of directed arcs. A Petri net is the triple $(P,Tr,F)$. Its marking $M\in\mathcal{B}(P)$ is a multiset of tokens and encodes the current state. Petri nets that have an initial marking $M_0$ and a final marking $M_{f}$, i.e., an initial and final state, are called accepting Petri nets. Furthermore, let $A\subseteq\mathcal{A}$ be a set of activities and $l_{Tr}:Tr\rightarrow A\,\cup\,\{\tau\}$ a function that associates to elements of $Tr$ either an activity of $A$ or the ``silent" label $\tau$, which represents unobservable activities that add legitimate behaviors not directly visible. A \emph{labeled accepting Petri net} is the tuple $(P,Tr,F, M_0, M_f,A,l_{Tr})$.
$\mathcal{N}$ denotes the universe of all labeled accepting Petri nets, hereafter referred to as Petri nets.
\end{definition}

The formal semantics of Petri nets include the ability to \textit{fire} a transition. Given the number of incoming arcs of a transition, if the current marking is such that there is at least a token in each place connected to the transition, it can fire, i.e., the transition consumes one token from each incoming place and generates one token for each outgoing arc. This firing dynamic enables the Petri net to evolve its state, allowing for the above-mentioned conformance checking and simulation. In addition, it is worth noting that there is a special class of Petri nets, namely workflow Petri nets, which have two specific places: a source place and a sink place. In these Petri nets, $M_0$ assigns a token to the source place. Any complete firing sequence moves the Petri net state from $M_0$ to $M_f$, which consists of at least a token in the sink place.

\begin{figure*}[!t]
\centering
\begin{subfigure}[b]{0.47\textwidth}
    \centering
    \includegraphics[width=\textwidth]{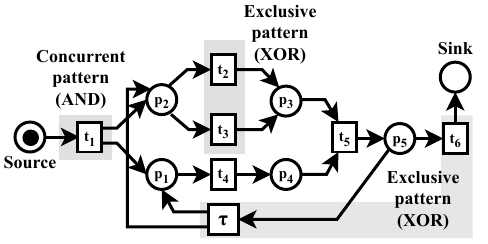}
    \caption{Petri net with grey-highlighted control-flow patterns.}
    \label{EXAMPLE_HIGHLIGHTED_PETRI_NET}
\end{subfigure}\hfill
\begin{subfigure}[b]{0.52\textwidth}
    \centering
    \includegraphics[width=\textwidth]{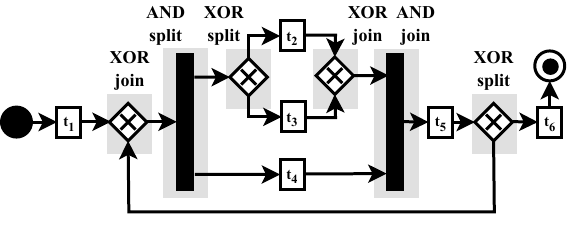}
    \caption{Trace-equivalent UML activity diagram.}
    \label{TRACE_EQUIVALENCE}
\end{subfigure}
\caption{Petri net and its trace-equivalent UML activity diagram.}
\label{PETRI_AND_AD}
\end{figure*}

Figure \ref{EXAMPLE_HIGHLIGHTED_PETRI_NET} depicts an example Petri net made of places \texttt{source}, \texttt{sink} and \texttt{p}$_{1,\dots,5}$, and transitions \texttt{t}$_{1\dots 6}$ and $\tau$. The figure highlights in grey the control-flow patterns that the Petri net captures, such as the concurrent pattern (i.e., the AND pattern) linked to \texttt{t}$_1$, \texttt{p}$_2$ and \texttt{p}$_3$. This pattern allows the concurrent execution of transition \texttt{t}$_4$ and either transition \texttt{t}$_2$ or \texttt{t}$_3$. A desirable property of Petri nets is ensuring that the final marking is always reachable and that no transitions or places become dead. This property, met by the example Petri net, guarantees a complete and token-free traversal of the net, ensuring every transition has a viable path from start to finish.

Notably, some transformation rules can also be applied to map semi-formal models to formal models. Figure \ref{TRACE_EQUIVALENCE} shows a trace-equivalent UML activity diagram of the Petri net in Figure \ref{EXAMPLE_HIGHLIGHTED_PETRI_NET}. The XOR and AND patterns are explicitly encoded by graphic elements of the UML notation. The UML activity diagram is trace-equivalent since any trace of the Petri net is also allowed on the UML diagram. For example, the trace $\langle t_1, t_2, t_4, t_5, t_3, t_4, t_5, t_6 \rangle$ can be executed on the UML activity diagram as follows: $t_1$ initiates two concurrent flows, with the upper path implementing an XOR choice that allows $t_2$ to execute, followed by $t_4$, and then $t_5$ after synchronization at the AND join. After this, the XOR split cycles back to the concurrent flow, where $t_3$ and $t_4$ execute, synchronize, and allow $t_5$ again, before the process concludes with $t_6$. This illustrates how the UML diagram faithfully represents both concurrency and exclusive choices encoded in the Petri net.

\subsection{Conformance checking and anomaly detection}
As the event log and Petri net are well-defined, the conformance checker can now implement a conformance checking technique to verify whether the traces of an event log follow the control-flow patterns prescribed by a reference Petri net. Such verification leads to a series of conformance-checking \textbf{diagnoses} that record two types of global and local information: the fitness and misalignments. These can be computed through alignment-based conformance checking, which first evaluates the alignments between traces and a reference Petri net. 

\begin{definition}[Alignment and moves]
\label{def:alignment}
Let us denote $\gg$ as the ``skipped" activity. Let $N\in\mathcal{N}$ be a reference Petri net and $\sigma\in \mathcal{A}^*$ a trace. $\sigma_L=\langle a_1,\dots,a_o\rangle\in (\mathcal{A}\cup\{\gg\})^*$ is a (finite) sequence of log moves related to $\sigma$ if and only if $\sigma_L\setminus \{\gg\}=\sigma$. Then, $\sigma_{N}=\langle b_1,\dots, b_o\rangle\in (\mathcal{A} \cup \{\tau\} \cup \{\gg\})^*$ is a (finite) sequence of model moves if and only if $\sigma_{N}\setminus\{\gg\}$ is a firing sequence of $N$, i.e. an allowed control flow. An \emph{alignment} $\gamma_{\sigma, N}$ is the two-row matrix
\begin{flalign*}
\gamma_{\sigma, N}&=
\begin{array}{c|c|c|c}
a_1 & a_2 & \cdots & a_o \\
\hline
b_1 & b_2 & \cdots & b_o
\end{array},&
\end{flalign*}
if for all $1 \leq i \leq o$, $(a_i, b_i) \neq (\nomove, \nomove)$, where $(a_i,b_i)$ is a \emph{move}. The upper row is the sequence of log moves (log sequence) and the bottom row is the sequence of model moves (model sequence). We define $|\gamma_{\sigma,N}|$ the \emph{alignment length}, i.e., the number of moves of the alignment.
\end{definition}

\begin{figure*}[!t]
\centering
\includegraphics[width=\textwidth]{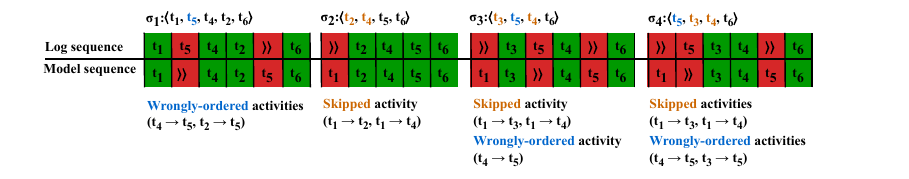}
\caption{The best alignments associated with four faulty traces against the Petri net in Figure \ref{EXAMPLE_HIGHLIGHTED_PETRI_NET}. The red-highlighted alignment parts indicate mismatches between the trace and the Petri net..}
\label{ALIGNMENTS}
\end{figure*}
Alignment-based conformance checking attempts to find the \emph{closest} (best) alignment between a reference Petri net and a trace. Figure \ref{ALIGNMENTS} shows the four best alignments against the Petri net in Figure \ref{EXAMPLE_HIGHLIGHTED_PETRI_NET} linked to four faulty traces $\sigma_{1,\dots,4}$. The red-highlighted alignment parts indicate mismatches between the trace and the Petri net due to violating some control-flow constraints. For example, $\sigma_1$ (leftmost trace) has wrongly-ordered activities, as \texttt{t}$_5$ precedes \texttt{t}$_4$ and \texttt{t}$_2$ instead of being executed after their appearance. This violation leads to the red-highlighted model-log sequence pairs, which can be recorded as additional diagnoses to use in the methodology's further steps.
Once the best alignments are found, the fitness can be computed.

\begin{definition}[Fitness]
\label{def:ab_fitness}
Let us denote $\Gamma_{\sigma, N}$ as the set of all alignments between a trace $\sigma\in\mathcal{A}^*$ and a Petri net $N\in\mathcal{N}$. Let $\delta$ be the unitary cost function for a pair of moves or an alignment. Finally, let $\gamma^w_{\sigma,N}\in\Gamma_{\sigma, N}$ be the worst-case alignment and $\gamma^*_{\sigma,N}\in\Gamma_{\sigma, N}$ the best-case alignment, i.e., the alignments with the least and most costs according to $\delta$, respectively.  The \emph{alignment-based fitness} for $\sigma$ is defined as
\begin{flalign*}
F_{\sigma,N}&=1-\frac{\delta(\gamma^*_{\sigma,N})}{\delta(\gamma^w_{\sigma,N})}.&
\end{flalign*}
Let $L\in\mathcal{B}(\mathcal{A^*})$ be an event log. The \emph{alignment-based fitness} for $L$ is defined as
\begin{flalign*}
F_{L,N}&=\frac{\sum_{\sigma\in L}F_{\sigma,N}}{|L|}.&
\end{flalign*}
\end{definition}

The misalignments are counters associated with each transition $tr\in Tr$ of the reference Petri net that record the number of mismatches between the log sequences and model sequences involving $tr$. For example, Figure \ref{ALIGNMENTS} outlines that there are 2 \texttt{t$_5$} mismatches in the best alignment of $\sigma_1$. A detailed description of how to calculate misalignments can be found in \cite{vitale2025cfad}. The misalignments and fitness constitute the diagnoses.

\begin{definition}[Diagnoses]
\label{def:cc_fitness}
Let $\mathcal{A}_\alpha\subseteq\mathcal{A}$ be the number of activities of the target software and $L\in\mathcal{B}(\mathcal{A_\alpha^*})$ be an event log of $k$ traces and $N\in\mathcal{N}$ a Petri net. The conformance checker collects the diagnoses $D\in\mathbb{R}^{k\times (\alpha+1)}$ as in Table \ref{cc_diagnoses} by alignment-based conformance checking of each trace $\sigma\in L$ against $N$.
\end{definition}
\begin{table}[!t]
\centering
\caption{Alignment-based conformance checking diagnoses obtained from replaying the traces $\sigma_1\dots\sigma_{|L_o|}$ of event log $L_o$ against a reference Petri net $N$.}
\label{cc_diagnoses}
\begin{tabular}{lllllll}
\hline
\textbf{Trace}            & $\boldsymbol{a_1}$   & $\boldsymbol{a_2}$   & $\cdots$ & $\boldsymbol{a_{\alpha-1}}$   & $\boldsymbol{a_{\alpha}}$   & $\boldsymbol{F_{\sigma}}$ \\ \hline
$\boldsymbol{\sigma_1}$   & $u_{a_1,\sigma_1}$   & $u_{a_2,\sigma_1}$   & $\cdots$ & $u_{a_{\alpha-1},\sigma_1}$   & $u_{a_{\alpha},\sigma_1}$   & $F_{\sigma_1}$            \\
$\boldsymbol{\sigma_2}$   & $u_{a_1,\sigma_2}$   & $u_{a_2,\sigma_2}$   & $\cdots$ & $u_{a_{\alpha-1},\sigma_2}$   & $u_{a_{\alpha},\sigma_2}$   & $F_{\sigma_2}$            \\
$\cdots$                  & $\cdots$             & $\cdots$             & $\cdots$ & $\cdots$                      & $\cdots$                    & $\cdots$                  \\
$\boldsymbol{\sigma_{k}}$ & $u_{a_1,\sigma_{k}}$ & $u_{a_2,\sigma_{k}}$ & $\cdots$ & $u_{a_{\alpha-1},\sigma_{k}}$ & $u_{a_{\alpha},\sigma_{k}}$ & $F_{\sigma_{k}}$          \\ \hline
\end{tabular}
\end{table}

These can be used by an anomaly detector to verify whether a trace is anomalous based on both its number of misalignments and the fitness value. Control-flow anomaly detection is performed by classifying each trace of the diagnoses.
\begin{definition}[Control-flow anomaly]
\label{def:cfa}
Let $D\in\mathbb{R}^{k\times (\alpha+1)}$ be the diagnoses collected by the conformance checker. Let $d_{\sigma}\in D$ be the diagnoses associated with a trace $\sigma$. $\sigma$ exhibits an \emph{control-flow anomaly} if the anomaly detector classifies $d_\sigma$ as anomalous.
\end{definition}

Given the structure of diagnoses, any unsupervised anomaly detection algorithm can be used. This class of algorithms identifies patterns or deviations without requiring labeled data, making them suitable for scenarios where normal and abnormal behaviors are not explicitly defined~\cite{zoppi2021unsupervisedad}. In this context, the algorithm interprets the diagnosis metrics, such as misalignment counts, fitness scores, and other derived features, as indicators of behavioral conformity. By modeling the distribution of these indicators across multiple traces, it can flag outliers that significantly deviate from the expected process behavior, thereby signaling potential anomalies in the system execution. The training of the anomaly detector will be detailed in the following section.

\section{The methodology for LLM-enabled software control-flow anomaly detection}
\label{sec:methodology}
\begin{figure*}[!t]
\centering
\includegraphics[width=\textwidth]{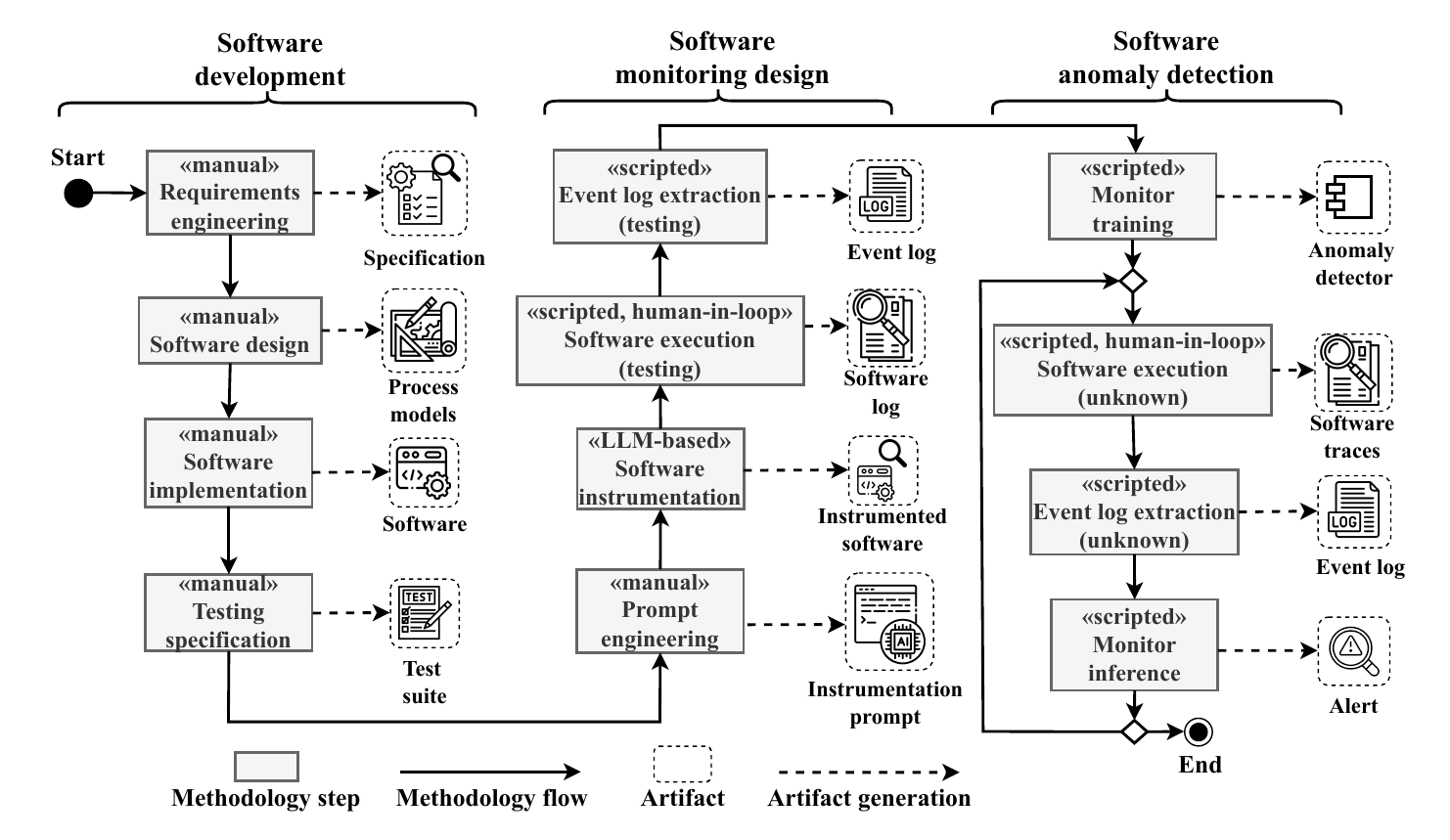}
\caption{The methodology for LLM-enabled software control-flow anomaly detection. The methodology is split into the software development, software monitoring design, and software anomaly detection phases. The software development phase follows the traditional development steps to specify, design, implement and test the software. The software monitoring design phase uses LLMs to link process models with the software implementation to enable the extraction of event logs during runtime testing. The software anomaly detection phase builds the anomaly detector using the logs collected during runtime testing for subsequent control-flow anomaly detection through conformance checking.}
\label{METHODOLOGY}
\end{figure*}

This section delves into the methodology that we devised to enable the use of the conformance checking-based monitor defined in Section \ref{sec:cc_based_monitor}. The methodology is depicted in Figure \ref{METHODOLOGY} and split into three phases: \textbf{software development}, \textbf{software monitoring design} and \textbf{software anomaly detection}. Each phase involves different steps, which can either be entirely \textit{manual}, \textit{LLM-based}, \textit{scripted} with \textit{human-in-loop}, or entirely scripted. Manual steps are performed by humans without relying on scripted code or LLMs; these are mostly linked to software development, where a high level of human supervision is required. The only LLM-based step is software instrumentation, which is triggered by a prompt. Software executions are both scripted and human-in-loop, as these usually involve some degree of human interaction. The rest of the steps are entirely automated through scripted code.

\subsection{Software development}
\subsubsection{Requirements engineering and software design} In the \textbf{requirements engineering step}, a requirements engineer analyzes the software system requirements and generates a System Requirements Specification (SRS) document reporting the software system's functional and non-functional aspects. The functional aspects regard the use cases that the software must implement, whereas non-functional aspects regard performance and dependability metrics. 
For example, the developers of the ERTMS/ETCS standard produced the ERTMS/ETCS SRS document (SUBSET-026). The dynamic aspects of the standard procedures are described in the document's chapter 5 (SUBSET-026-5), including the SoM scenario. 

As will be shown in Section \ref{sec:case_study}, the SRS is mostly written in natural language and can be subject to interpretation ambiguity. This ambiguity can initially be solved with the development of formal models in the subsequent \textbf{software design} step. The SRS is used to drive this step and design of the software architecture, which is captured by several diagrams describing static and dynamic aspects. Static aspects involve the description of components' responsibilities and their interconnection. Dynamic aspects involve the procedures that components execute and how the procedures relate to each other. 

Regarding the dynamic aspects, software design typically relies on high-level modeling languages such as UML activity diagrams and the Business Process Modeling Notation (BPMN). As illustrated in Figure~\ref{PETRI_AND_AD}, the labeled accepting Petri net (Definition~\ref{def:petri_net}) provides a formal bridge from these semi-formal languages to a rigorous execution framework. In this mapping, activities correspond to labeled transitions, while control-flow constructs such as loops and choices are represented through networks of arcs, places, and silent ($\tau$) transitions.

Crucially, unlike plain Petri nets, the accepting property (defined by an initial marking $M_0$ and a final marking $M_f$) allows capturing the execution semantics of software design: high-level notations such as UML and BPMN explicitly specify start and end points, which can be directly mapped to the source place (containing the initial marking $M_0$) and the sink place (which should exclusively contain a token upon trace completion, corresponding to the final marking $M_f$).

Moreover, the use of labeled accepting Petri nets also enables the alignment-based conformance checking logic of the proposed monitor (Section \ref{sec:cc_based_monitor}). In fact, during this process, $\tau$-transitions are handled by assigning them a move cost of zero, whereas the alignment with the expected sequencing of labeled transitions enables quantifying the degree of conformance to prescriptive behavior. The entire process is also strongly supported by the existing tools and libraries, such as \texttt{pm4py}\footnote{\url{https://processintelligence.solutions/pm4py/}} and \texttt{ProM}\footnote{\url{https://promtools.org/}}. It is worth noting that partial or interrupted runs can penalize the resulting diagnoses, marking more misalignments. This is due to the traces resulting from partial or interrupted runs reaching an invalid final marking. In this case, prefix-alignments should be adopted~\cite{schuster2020prefixalignment}.

During software design, the development team can start the verification and validation (V\&V) process through \textit{design-time testing} to ensure that the design artifacts meet the requirements. Comprehensive V\&V provides the evidence needed to support safety claims and regulatory approval, especially in the context of software certification for compliance with relevant standards such as EN 50128, IEC 61508, or ISO 26262. In this case, the use of Petri nets is even more advantageous due to their formal syntax and semantics, which allow verifying whether the dynamic description leads to deadlocks, unwanted control-flow relationships, or simply invalidates the requirements. Despite its importance, design-time testing is not sufficient to prove the correctness of the software; more testing is needed in the subsequent phases.

\subsubsection{Software implementation and testing specification} 
Once the software has been thoroughly documented and its behavior, structure, and interactions have been described through process models, the next step is \textbf{software implementation}. At this stage, developers translate conceptual representations into executable code,
and V\&V progresses by employing white-box testing and black-box testing to ensure that the software implementation conforms to its specifications and fulfills intended requirements~\cite{ISO29119_2022}. On the one hand, white-box testing evaluates the software correctness by inspecting its structure either statically or dynamically. It typically involves developing test suites to cover the majority of statements, paths, and/or conditions. On the other hand, black-box testing evaluates the software correctness by assessing the output conditioned by a given input. It also involves developing test suites, although they are aimed at defining what the correct output is. Through white-box and black-box testing, the \textbf{testing specification} step generates a test suite.

However, despite the completeness of the test suite, runtime conditions may give rise to so-called “unknown unknowns'', which are situations that were not anticipated before the software system is deployed~\cite{sinclair2025chasing}. These situations can occur due to \textit{residual faults}, which are defects that escape test suites~\cite{natella2013sfi}.
Detecting the activation of residual faults is important for maintaining system safety and reliability, as these faults lead to latent errors that may eventually cause system failures~\cite{carnevali2025faultflow}. One effective approach is to employ a runtime monitor that tracks software behavior as it unfolds, enabling the system to respond to unexpected or unpredictable conditions \cite{cinque2016characterizingdirectmonitoring}.

\subsection{Software monitoring design}
\subsubsection{Prompt engineering and software instrumentation}
In our methodology, the runtime monitor requires checking deviations from the prescriptive design-time models by analyzing software logs obtained from software executions. Among the different types of instrumentation techniques, source-code instrumentation is particularly promising, as it enables collecting semantically meaningful events tied to the software’s structural representation. 
However, since software implementations often deviate significantly from high-level descriptions, bridging low-level software constructs with design-time models remains challenging. To address this, we propose using LLMs to assist in source-code instrumentation, leveraging their ability to 1) encode domain knowledge and 2) infer semantic structure from code.

While LLMs are powerful tools, it is widely acknowledged that their output is strongly dependent on the prompts used. This gave rise to \textbf{prompt engineering}, an activity concerned with designing prompt structures that effectively guide the model toward producing accurate, relevant, and contextually appropriate responses. The following is the logging prompt that we will use to instrument the code. Thus, in the \textbf{software instrumentation} step, the LLM is provided with the logging prompt, the reference process model, and the software. By acknowledging the dynamics described in the process model and mapping them to the software constructs, the LLM is capable of bridging the high-level model with the low-level implementation, providing an instrumented software that generates quality software logs.

\subsubsection{Software execution and event log extraction}

The test suite generated during software development can be used to stimulate the \textbf{software execution}. Such \textit{testing} yields a software log, which contains the different events logged according to the instrumentation rules. However, while the LLM has used high-level descriptions of design-time models, the software log may still have spurious information that is not relevant to the high-level reference process model. Accordingly, \textbf{event log extraction} filters out spurious information and retains only the sequence of activities mapped to the process model activities, resulting in an event log compliant with Definition \ref{def:event_log}.

\subsection{Software anomaly detection}
\subsubsection{Monitor training} 
To build the monitor, the event log extracted during software monitoring design is used in the \textbf{monitor training} step to train the anomaly detector to classify control-flow anomalies. 
\begin{definition}[Monitor training]
Let $\Delta$ denote the universe of binary classifiers. Let $L\in\mathcal{B}(\mathcal{A}_\alpha^*)$ be an event log of $k$ traces obtained from executing $k$ tests against the instrumented software. Let $N\in\mathcal{N}$ be a Petri net described during the software development phase. Let $D\in\mathbb{R}^{k\times (\alpha+1)}$ be the diagnoses obtained by the conformance checker by aligning the traces of $L$ against $N$. \emph{Monitor training} is the function $\eta:\mathbb{R}^{k\times o}\rightarrow \Delta$, such that $\eta(D)=\delta$, where $\delta$ is the anomaly detector.
\end{definition}
Through the anomaly detector, the conformance checking-based monitor is capable of classifying a new piece of data as either anomalous or normal. Many unsupervised machine learning algorithms can implement $\eta$ and build a monitor. Two common approaches are clustering and dimensionality reduction \cite{zoppi2021unsupervisedad}. Clustering involves grouping similar data points using a distance metric, such as the Euclidean distance. A threshold on the distance from the closest cluster can be set so that any new piece of data exceeding such distance is deemed anomalous. Dimensionality reduction involves building a new lower-dimensional reference system on which data are projected. Reconstructing data from the lower-dimensional reference system to the original one leads to reconstruction error. A threshold on the reconstruction error can be set so that any new piece of data exceeding such error is deemed anomalous.

\subsubsection{Software execution, event log extraction, and monitor inference} When unknown software executions occur as the users interact with the software, new traces are collected using the event log extraction prompt above. To verify whether the software executed as intended, \textbf{monitor inference} handles the traces through conformance checking and evaluates whether it exhibits a control-flow anomaly according to Definition \ref{def:cfa}.

\section{Case study}
\label{sec:case_study}
\begin{figure}[!t]
\centering
\includegraphics[width=\columnwidth]{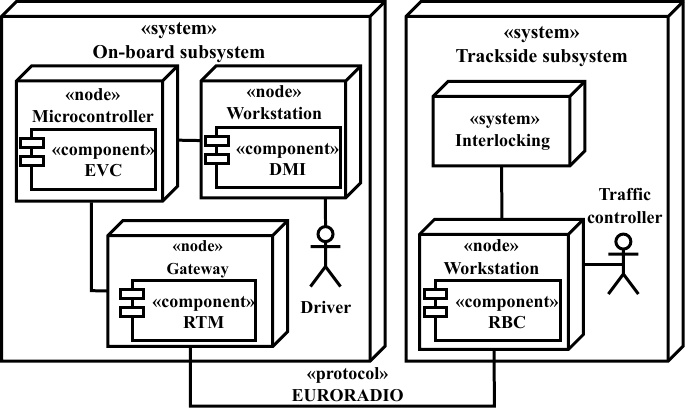}
\caption{Deployment diagram of ERTMS/ETCS railway systems, split into the on-board subsystem and trackside subsystem.}
\label{ERTMS_DEPLOYMENT}
\end{figure}
In this section, we demonstrate the application of the proposed methodology in Figure \ref{METHODOLOGY} to the SoM procedure of the ERTMS/ETCS standard.

\subsection{Software development}
The ERTMS/ETCS SRS is mostly written in natural language and can be subject to interpretation ambiguity between different manufacturers of system components. In fact, extensive field integration testing involving different companies is usually required, which, however, cannot exclude interaction anomalies in unspecified situations and edge cases, and that can introduce additional uncertainties, which can be possibly managed using the approach described in this paper. The following are a few excerpts from SUBSET-025-5:
\begin{excerpt}[Excerpts from SUBSET-026-5, §5.4.2\\Status of data stored in the ERTMS/ETCS on-board equipment]
\textbf{§5.4.2.1:} At the beginning of the Start of Mission procedure, the data required may be in one of 
three states: 
\begin{enumerate}[label=\alph*), nosep]
  \item \textbf{Valid} — the stored value is known to be correct.
  \item \textbf{Invalid} — the stored value may be wrong.
  \item \textbf{Unknown} — no stored value is available.
\end{enumerate}

\smallskip

\textbf{§5.4.2.2:} This refers to the following data: Driver ID, ERTMS/ETCS level, RBC contact information, Train Data, Train Running Number, Train Position (see §3.6.1.3).

\smallskip

\textbf{§5.4.2.3:} Note 1: The status of data in relation to the previous and the actual mode is described in 
chapter 4, section ``What happens to stored information when entering a mode''. 

\textbf{§5.4.2.4:} Note 2: The change of status of data in course of the procedure is shown in the table in §5.4.3.3. 
\end{excerpt}
\noindent Although these informal descriptions of the requirements are quite structured and are attached with detailed tables and, at times, state diagrams, they still need to be translated into artifacts that document the progressive refinement of natural language into more rigorous and possibly formal models. 

\begin{figure*}[!t]
\centering
\includegraphics[width=\textwidth]{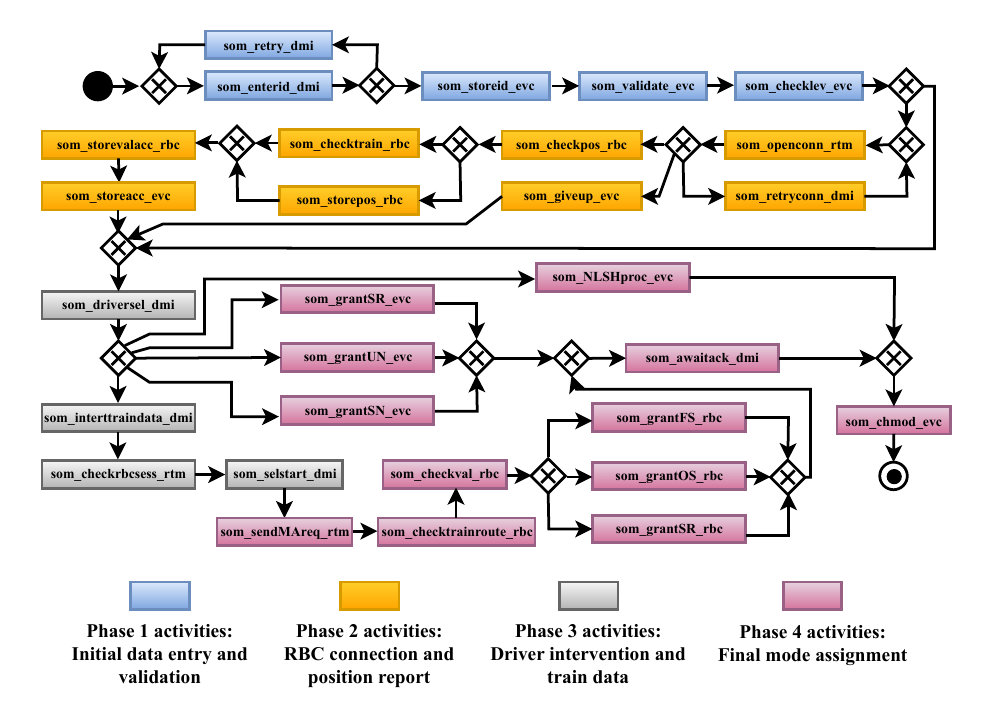}
\caption{High-level UML activity diagram describing the SoM procedure, split into four phases: initial data entry and validation; RBC connection and position report; driver intervention and train data; and final mode assignment.}
\label{SOM_PROCEDURE}
\end{figure*}
We have designed two UML diagrams representing simplified views of the essential static and dynamic aspects of ERTMS/ETCS railway systems. Figure \ref{ERTMS_DEPLOYMENT} illustrates the core components of the on-board and trackside subsystems. This diagram captures the deployment of ERTMS/ETCS components—EVC, DMI, RTM, RBC, and interlocking—as specified in SUBSET-026-2, and depicts the interconnection of the on-board and trackside subsystems through the EURORADIO protocol.

Building on this structure and interpreting the requirements of SUBSET-026-5, §5.4, we have developed a UML activity diagram representing the high-level Start of Mission (SoM) control flow, shown in Figure \ref{SOM_PROCEDURE}. The activity diagram abstracts implementation details such as specific message exchanges, data structures, and state-machine transitions, focusing instead on the logical sequencing of activities and decision points.

To provide clarity, the UML activities have been organized into four categories, corresponding to the essential phases of SoM: 1) initial data entry and validation; 2) RBC connection and position report; 3) driver intervention and train data; and 4) final mode assignment. Each activity is labeled using a repeatable and interpretable naming scheme: \\\texttt{som\_<action\_name>\_<component>}, where \texttt{<component>} corresponds to one of the components depicted in Figure \ref{ERTMS_DEPLOYMENT}, and \texttt{<action\_name>} describes the specific action performed. This systematic labeling ensures repeatability and facilitates understanding across the SoM workflow.

\begin{lstlisting}[
    style=pythonstyle, 
    caption={Code snippet of the prototype software implementing phase 1 of the SoM procedure.},
    label={code:phase1},
    float=t! 
]
def _phase_1_get_initial_data(self):
    if not (self.current_mode == Mode.STAND_BY and self.desk_open):
        return None 
    state = self._procedure_s1_driver_id_entry()
    if state == 'D2':
        state = self._procedure_d2_check_pos_level()
    if state == 'S2':
        state = self._procedure_s2_level_entry()
    if state == 'D3':
        state = self._procedure_d3_check_level_valid()
        if state == 'D7': 
            return 'S3' 
    return state 

def _procedure_s1_driver_id_entry(self):
    if self.driver_id_status == DataStatus.UNKNOWN:
        self.driver_id = self._simulate_driver_action("Please enter Driver ID:")
    elif self.driver_id_status == DataStatus.INVALID:
        self.driver_id = self._simulate_driver_action("Please revalidate/re-enter Driver ID:")
    self.driver_id_status = DataStatus.VALID
    return 'D2'
\end{lstlisting}

As explained in the previous sections, transformation rules can be applied to obtain a trace-equivalent Petri net from the UML activity diagram. This is noteworthy since conformance checking algorithms are typically suited for the Petri net formalism. 

The specification and the diagrams have guided the implementation of an SoM prototype in Python, which adds more details than the essential behavior of the high-level diagrams shown earlier. Listing~\ref{code:phase1} shows a code snippet of the prototype software implementing phase 1 of the SoM procedure, including the subroutine that handles the acquisition of the driver ID from the DMI.

\begin{lstlisting}[
    style=pythonstyle, 
    caption={Code snippet of the test to obtain full supervision.},
    label={code:test-l2-fs},
    float=t!
]
def test_path_l2_success_grant_fs(setup_test_logger):
    script = [
        "DRIVER_007", '2', True, "session_open", "validate_position",
        'TD', "DATA", "TRN", "ack", "Start", "ma_fs"
    ]
    sim = ERTMSOnBoardSystem(logger=setup_test_logger, controller=MockController(script))
    sim.run_start_of_mission()
    assert sim.current_mode == Mode.FULL_SUPERVISION
\end{lstlisting}
Next, we developed a comprehensive test suite that instantiated 50 control-flow paths by stimulating the prototype with different inputs. The test suite must cover a wide range of possibilities, as the SoM procedure may instantiate many different control flows. Listing~\ref{code:test-l2-fs} shows one of the tests in the test suite aimed at forcing the software to provide full supervision to the train.

\subsection{Software monitoring design}
\begin{lstlisting}[
    style=pythonstyle, 
    caption={Code snippet of the \_procedure\_s1\_driver\_id\_entry function instrumented by Gemini Pro 3.0.},
    label={code:proc-s1},
    float=!t
]
def _procedure_s1_driver_id_entry(self):
    if self.driver_id_status == DataStatus.UNKNOWN:
        self.logger.info("som_enterid_dmi")
        self.driver_id = self._simulate_driver_action("Please enter Driver ID:")
    elif self.driver_id_status == DataStatus.INVALID:
        self.logger.info("som_retry_dmi")
        self.driver_id = self._simulate_driver_action("Please revalidate/re-enter Driver ID:")
    
    self.logger.info("som_storeid_evc")
    self.driver_id_status = DataStatus.VALID
    
    self.logger.info("som_validate_evc")
    return 'D2'
\end{lstlisting}
Having developed both the software and its test suite, the instrumentation can now be performed through an LLM. Specifically, we have used the software instrumentation prompt seen in Section \ref{sec:methodology} and provided LLMs with both the prototype software and the process model in XML format. With both domain-specific knowledge and full access to the code internals, LLMs are able to interpret code structure and place logging instructions where they identify the best fit. For example, Listing \ref{code:proc-s1} shows the instrumentation of one of the phase 1 procedures shown in Listing \ref{code:phase1} by Gemini Pro 3.0, one of the LLMs we used in our experiments. The LLM has successfully identified the correct code locations where the \texttt{som\_enterid\_dmi}, \texttt{som\_retry\_dmi}, \texttt{som\_storeid\_evc}, and \texttt{som\_validate\_evc} activities had to be logged.

\subsection{Software anomaly detection}
The event log resulting from the software execution with the test suite can now be used for monitor training. This is necessary because although LLMs allow obtaining quality event logs, they may introduce some noise, such as missing or duplicated activities. While these discrepancies may indicate anomalies, they can be treated as source-code instrumentation noise. Thus, by allowing some degree of noise, the monitor can become more effective at finding true positives while reducing the amount of false positives thanks to this tolerance.

Since the only available traces are those obtained through the test suite, the event log only contains normal traces. After performing conformance checking against the reference Petri net, unsupervised, one-class algorithms can be used to learn how to characterize normal behavior from the diagnoses of Definition \ref{def:cc_fitness}. In the next section, we will show the application of several unsupervised, one-class machine learning techniques to the diagnoses generated through our methodology.
\section{Evaluation}
\label{sec:evaluation}
The evaluation aims to substantiate the answers to the research questions with metrics that quantify source-code instrumentation quality (RQ1) and control-flow anomaly detection effectiveness (RQ2). The software for the experiments is implemented using Python and was run on a Windows 11 machine with an Intel® Core™ i9-11900K CPU @ 3.50GHz and 32GB of RAM. The software uses machine learning and process mining libraries, such as \texttt{scikit-learn} and \texttt{pm4py}. In addition, it uses the APIs of several LLMs to query the models with the prompts shown in Section \ref{sec:methodology}. The software is available online on GitHub\footnote{\url{https://github.com/francescovitale/pm_software_monitor}}.

\subsection{RQ1: Source-code instrumentation}

\subsubsection{Experimental setup}
\paragraph{LLM model set} In order to provide a diversified analysis of the source-code instrumentation quality obtained by LLMs, we are evaluating the output of different state-of-the-art models. In particular, we focus on LLMs that exhibit implicit reasoning capabilities. Unlike explicit Chain-of-Thought (CoT) approaches that rely on external triggers, implicit reasoning allows models to decompose complex problems into intermediate steps within their internal hidden states or latent representations, without requiring specific step-by-step prompting~\cite{deng2023implicitchainthoughtreasoning, wang2024chainofthoughtreasoningprompting}.
We have considered two state-of-the-art LLMs that have been reported to have reasoning capabilities: Gemini 3.0 Pro (gemini-3.0-pro) \cite{GoogleDeepMind2025Gemini3} and Claude Sonnet 4.5 (claude-sonnet-4.5) \cite{claudeSonnet45}. Notably, the Gemini and Sonnet families have shown remarkable performance in software engineering tasks, as reported in the SWE-bench results~\cite{jimenez2024swebench}. 
In addition, we have considered Gemini 3.0 Flash (gemini-3.0-flash) and Claude Haiku 4.5 (claude-haiku-4.5), which offer shallower reasoning capabilities that are expected to influence the end results. 
Finally, it is worth noting that these LLMs expose several parameters that influence the determinism of their outputs, including temperature, top-p, and top-k sampling~\cite{li2025temperatureimpact}.
In our experiments, all models were evaluated using the default inference parameters specified by their respective APIs (e.g., a temperature of 1.0). While such defaults introduce controlled non-determinism, they are designed to balance coherence and diversity, enabling exploratory generation rather than strict greedy decoding~\cite{renze2024samplingtemperatureeffect}.
Adhering to provider-defined default settings ensures consistency across models and reflects their intended real-world usage, thereby supporting fair and comparable evaluation.

\paragraph{Instrumentation prompt}
Each LLM is fed with a prompt to guide the software instrumentation. The prompt includes input artifacts, assigns a role to the LLM, provides the context, specifies point-to-point actions, and establishes the output type. The instrumentation prompt used for each LLM is the following.
\begin{prompt}[Instrumentation prompt]
\textbf{Input:} The SoM source code, the test suite, and the XML version of the process model.\\
\textbf{Role:} You are a software developer responsible for instrumenting an existing software system.\\
\textbf{Context:} You are instrumenting the source code of an ERTMS/ETCS-compliant prototype of the Start-of-Mission procedure. The prototype simulates the different control flows of the procedure documented in the standard with varying degrees of randomness. The prototype needs to be instrumented. You have access to the process model that guided the development process. You have access to the test suite that will be executed with pytest.
\\
\textbf{Action:} Instrument the code through these guidelines:
\begin{enumerate}[label=\arabic*), nosep]
    \item Do not modify the code logic.
    \item Instrument the code following the control flow of the process model and its activities.
    \item Before instrumenting, extract and list all activity names from the XML file.
    \item Only log the process model activities without adding any additional text. 
    \item Only use activity labels that appear in this exact list - do not invent or infer any labels.
    \item Make the instrumented code callable from the test suite. Only ensure that it is callable, do not perform any further fixes.
    \item Do not provide any further output than the one specified below.
\end{enumerate}
\textbf{Output:} The instrumented software.
\end{prompt}
\noindent The actions of the prompt guide the LLM throughout the process. In particular: action 1) ensures that the system under test's logic is not invalidated; action 2) constrains the instrumentation process to force the alignment with the process model's description; action 3) explicitly tells the LLM to extract the list of activities within the process model to have a matching between the software execution and the process model; action 4) ensures that the log does not add noise to the monitored software; action 5) shields against possible hallucinations in recording the process model activities; action 6) ensures a matching between the test suite and the code for logging its execution without modifying neither the code nor the test suite; and action 7) avoids post-processing actions that are useless for the task at hand.

\paragraph{Metrics}The quality of source-code instrumentation is evaluated using the following metrics. By comparing the event log obtained from the software execution using the test suite with the reference Petri net, we obtain the fitness. The higher the fitness, the higher the adherence of the control flow to the design-time description. Regarding event log-specific metrics, we consider the average trace length and the number of trace variants. Finally, we consider a key metric: the coverage. Let $L$ and $N$ be the event log resulting from the execution of the test suite and $N$ the reference Petri net, we define the control-flow coverage as follows:
\begin{flalign*}
\text{Coverage} &= 1-\frac{\#misalignments}{\sum_{\sigma\in L}|\gamma^*_{\sigma,N}|}.&
\end{flalign*}
The coverage indicates how aligned the source-code instrumentation is to the Petri net.

\paragraph{Data generation and experiment runs}
The four LLMs are independently stimulated with the instrumentation prompt for five runs each, resulting in five independent experiment executions per model. This is to account for the variability introduced by the non-zero temperature. For each run, the model context is fully reset, i.e., no conversational history or prior outputs are retained across runs. Once the instrumented version of the SoM source code is obtained, the test suite is executed and the software logs collected. Next, alignment-based conformance checking is performed for each trace of the software log, capturing both the fitness and coverage. In addition to the intra-LLM validation of the anomaly detection quality, we are including an inter-LLM validation that aims at verifying whether the quality of the instrumentation performed by different LLMs impacts anomaly detection. To this aim, we evaluate whether the conformance checking-based monitor obtained with the best-performing LLM is able to correctly identify normal and anomalous traces collected by the less-performing LLMs.

\begin{table*}[!t]
\centering
\caption{The fitness, number of variants, average trace length and coverage related to the source-code instrumentation performed by each LLM. The bold values indicate the best-performing LLM according to the control-flow coverage percentage.}
\label{tab:INSTR_QUALITY}
\begin{tabular}{llcccc}
\hline
\textbf{Reasoning} & \textbf{Model} & \multicolumn{1}{l}{\textbf{Fitness (\%)}} & \multicolumn{1}{l}{\textbf{Coverage (\%)}} & \multicolumn{1}{l}{\textbf{Trace Length}} & \multicolumn{1}{l}{\textbf{Variants}} \\ \hline
\multirow{2}{*}{Deep} & \textbf{claude-sonnet-4.5} & \textbf{86.617 ± 2.366} & \textbf{82.849 ± 2.636} & \textbf{14 ± 1} & \textbf{22 ± 1} \\
 & gemini-3.0-pro & 81.674 ± 3.060 & 77.579 ± 3.292 & 12 ± 1 & 23 ± 1 \\ \hline
\multirow{2}{*}{Shallow} & gemini-3.0-flash & 71.426 ± 7.335 & 66.823 ± 7.042 & 15 ± 1 & 25 ± 0 \\
 & claude-haiku-4.5 & 43.759 ± 15.410 & 46.391 ± 10.679 & 8 ± 2 & 20 ± 3 \\ \hline
\multicolumn{6}{l}{\textbf{Kruskal-Wallis test p-values: $p_{Fitness}<0.001$, $p_{Coverage}<0.001$}} \\ \hline
\end{tabular}
\end{table*}
\begin{figure*}[!t]
\centering
\includegraphics[width=\textwidth]{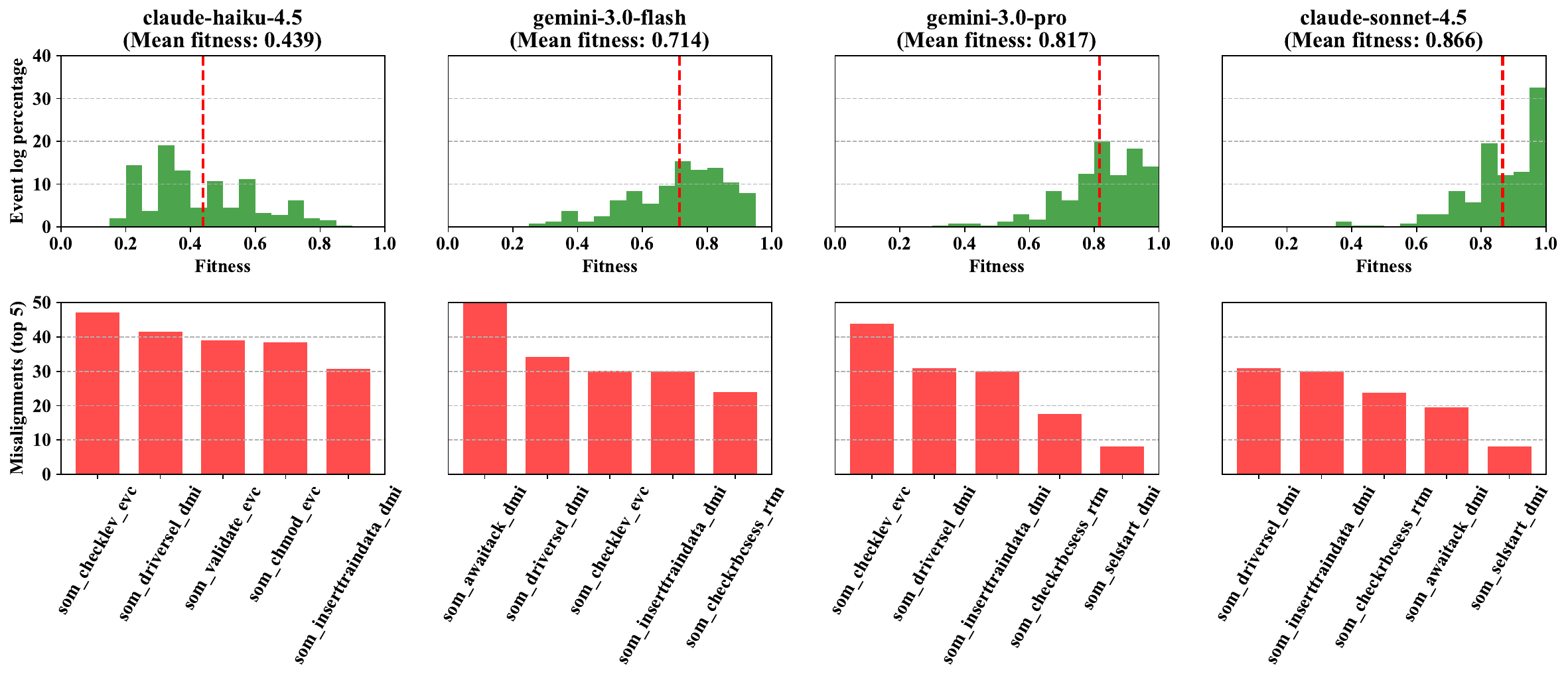}
\caption{The fitness value distributions of the event logs obtained for each LLM, and the corresponding top-5 per-activity misalignments.}
\label{FITNESS_MISALIGNMENTS}
\end{figure*}

\subsubsection{Results}
Table \ref{tab:INSTR_QUALITY} reports the results obtained for each LLM. The LLMs are further distinguished by their reasoning capabilities, placing them at either deep or shallow reasoning. As expected, the two best-performing LLMs with respect to the coverage metric are those with deeper reasoning, namely gemini-3.0-pro and claude-sonnet-4.5, which achieve 77.579\% and 82.849\% coverage while maintaining high fitness values (81.674\% and 86.617\%). This indicates that these models provided a high-quality instrumentation that aligns with the model (see Definition \ref{def:ab_fitness}) and covers most of its control-flow paths. The two shallower LLMs, gemini-3.0-flash and claude-haiku-4.5, could not achieve the same performance, dropping fitness (71.426\% and 43.759\%) and coverage (66.823\% and 46.391\%). These differences in performance are statistically significant, as highlighted by the p-values obtained through the Kruskal-Wallis statistical test. 

Figure \ref{FITNESS_MISALIGNMENTS} shows the fitness values distributions (top) of the traces of the event logs linked to the LLMs. The best fitness distribution is achieved by claude-sonnet-4.5, for which more than 30.0\% of traces achieve a fitness of up to $\approx 1$. This can also be visualized in the bar plots below, which show the top-5 misaligned activities by counting their misalignments (see Definition \ref{def:alignment}). The number of misalignments for each eactivity increases as the fitness distribution gets lower, highlighting the correlation between the global indicator and the local diagnoses.
It is worth noting that the best-performing LLMs, claude-sonnet-4.5 and gemini-3.0-pro, have similar misaligned activities, which also suggests that there may be some systematic errors in the implementation of the source code.

\subsection{RQ2: Control-flow anomaly detection}
\subsubsection{LLM selection and cross-validation}
We are going to evaluate and validate the ability of identifying control-flow anomalies with the proposed conformance checking-based monitor by exploiting the instrumentation performed by the LLMs as follows. First, we select the software logs from claude-sonnet-4.5 and internally validate the anomaly detection performance of the conformance checking-based monitor. Next, we perform a cross-validation of the conformance checking-based monitor of claude-sonnet-4.5 with the software logs generated from the other LLMs.

\subsubsection{Fault injection}
\paragraph{Fault types}To evaluate the capability of LLM-enabled conformance checking-based monitors, we inject three types of control-flow anomalies:
\begin{itemize}
    \item \textbf{Missing Activities (MA)}: The control flow is corrupted by removing an activity from the trace.
    \item \textbf{Unknown Activities (UA)}: The control flow adds activities that were not included in the model.
    \item \textbf{Wrongly-Ordered Activities (WOA)}: The control flow swaps the order of activities, opposed to the model prescriptions.
\end{itemize}
These three anomalies can be linked to the software fault taxonomy of Duraes and Madeira \cite{duraes2006softwarefaulemulation, natella2013sfi}. 
Specifically, the MA anomaly type corresponds to their \textit{missing/wrong construct} category, which the authors identified as the dominant type of software bug. For example, a missed \texttt{if} construct could skip the execution of an activity, or a wrong construct could not cover a condition in which an activity is expected to be executed.
Similarly, the UA anomaly type maps to their \textit{extraneous construct} category, representing surplus code. For example, redundant code could be wrongfully inserted by a programmer, causing the duplication of an activity.
Finally, the WOA anomaly type is a clear example of their \textit{wrong construct} category, which includes defects where the code is wrongly coded or ill-formed. For example, a statement could be misplaced within the control flow, causing an activity to be executed before its prerequisites are met, or sequential activities to be executed in the reverse order.

\paragraph{Injection process}The injection process is as follows. Given an event log $L\in\mathcal{B}(\mathcal{A}_\alpha^*)$ generated by simulating the model in Figure \ref{SOM_PROCEDURE}, we inject control-flow anomalies according to a Poissonian mechanism governed by a rate parameter \( \lambda > 0 \). For each trace $\sigma\in L$, the number of injected anomalies \( K \) is drawn from a Poisson distribution $K \sim \mathrm{Poisson}(\lambda)$, so that $P(K = k) = \frac{e^{-\lambda} \lambda^k}{k!}, k = 0, 1, 2, \dots$. The value of $K$ determines how many modifications are applied to trace $\sigma$. Let $AT \in \{\mathrm{MA}, \mathrm{WOA}, \mathrm{UA}\}$ denote the anomaly type, for which MA involves randomly deleting $K$ activities from $\sigma$, WOA involves random $K$ swaps between pairs of activities in $\sigma$, and UA involves randomly inserting $K$ new activities drawn uniformly from an unknown pool of activities at random positions in $\sigma$. $\sigma$ is transformed into an anomalous version $\tilde{\sigma}$ by applying the corresponding number and type of modifications: $\tilde{\sigma} = f(\sigma, AT, K)$, where \( f(\cdot) \) denotes the stochastic transformation induced by the selected anomaly mechanism. In this way, the Poisson parameter \( \lambda \) controls the expected number of anomalies per trace, providing a flexible and statistically grounded mechanism for simulating heterogeneous anomalous behavior in event logs.

\begin{table*}[!t]
\centering
\caption{The anomaly detection metrics associated with each technique and anomaly type using the software logs generated through claude-sonnet-4.5. The bold figures indicate the best-performing technique for the selected metric per anomaly type.}
\label{tab:AD_QUALITY}
\begin{tabular}{lllllll}
\hline
\textbf{Anomaly} & \textbf{Technique} & \textbf{Accuracy (\%)} & \textbf{Recall (\%)} & \textbf{Precision (\%)} & \textbf{F1-score (\%)} & \textbf{AUC (\%)} \\ \hline
\multirow{3}{*}{MA} & FT & 46.757 ± 9.767 & 32.400 ± 19.815 & 74.198 ± 7.291 & 41.606 ± 20.120 & 61.583 ± 9.605 \\
 & DBSCAN & 75.135 ± 3.686 & \textbf{99.600 ± 0.800} & 73.376 ± 3.064 & 84.456 ± 1.902 & 80.050 ± 6.352 \\
 & AE & \textbf{88.919 ± 2.620} & 92.000 ± 3.578 & \textbf{91.703 ± 2.346} & \textbf{91.798 ± 2.016} & \textbf{91.358 ± 4.411} \\
\hline
\multirow{3}{*}{WOA} & FT & 72.703 ± 8.126 & 70.800 ± 18.787 & 87.966 ± 5.432 & 76.365 ± 10.541 & 84.767 ± 2.470 \\
 & DBSCAN & 83.784 ± 6.837 & \textbf{98.000 ± 3.098} & 82.741 ± 8.613 & 89.347 ± 3.976 & 92.717 ± 4.283 \\
 & AE & \textbf{92.162 ± 1.576} & 95.200 ± 2.400 & \textbf{93.408 ± 2.152} & \textbf{94.255 ± 1.150} & \textbf{95.033 ± 3.015} \\
\hline
\multirow{3}{*}{UA} & FT & 39.865 ± 12.550 & 24.500 ± 24.551 & 50.877 ± 19.848 & 29.820 ± 25.549 & 56.208 ± 7.099 \\
 & DBSCAN & 81.081 ± 4.441 & 94.000 ± 8.000 & 82.407 ± 8.245 & 87.120 ± 2.530 & 90.783 ± 5.826 \\
 & AE & \textbf{92.973 ± 2.886} & \textbf{97.200 ± 2.040} & \textbf{92.777 ± 2.413} & \textbf{94.928 ± 2.067} & \textbf{94.617 ± 2.399} \\
\hline
\multirow{3}{*}{ALL} & FT & 45.862 ± 15.072 & 40.933 ± 19.811 & 92.437 ± 2.619 & 53.785 ± 18.496 & 66.989 ± 6.036 \\
 & DBSCAN & 89.425 ± 1.568 & \textbf{99.867 ± 0.267} & 89.195 ± 1.537 & 94.222 ± 0.801 & 89.456 ± 3.655 \\
 & AE & \textbf{93.103 ± 1.149} & 94.933 ± 0.800 & \textbf{97.009 ± 0.908} & \textbf{95.957 ± 0.673} & \textbf{93.669 ± 1.610} \\
\hline
\end{tabular}
\end{table*}
\subsubsection{Experimental setup}

\paragraph{Techniques} The selected conformance checking-based techniques are trained according to the framework in \cite{vitale2025cfad}, where the normal event log is split into three sublogs: the training, validation and test event logs. Diagnoses are extracted for each event log. The training diagnoses are used to train two unsupervised machine learning techniques: the density-based spatial clustering of applications with noise (DBSCAN), which deals with noisy, high-dimensional and arbitrarily distributed data \cite{schubert2017dbscan}, and the autoencoder (AE), which is a neural network-based approach for dimensionality reduction that involves encoding and decoding the input data with the aim of minimizing the reconstruction error \cite{sakurada2014adeutoencoders}. The validation diagnoses are used in two slightly different ways. For DBSCAN, a threshold is computed based on the distances of the validation data from the training clusters, whereas for AE, a threshold is built on the reconstruction error of validation data. The anomalous event logs are handled similarly and classified according to the threshold used. We also consider the baseline Fitness Thresholding (FT) technique, which is solely based on setting a threshold on the fitness metric. This is the most intuitive approach for conformance checking-based anomaly detection, although it is not as effective. 

\paragraph{Metrics}
We use the widely recognized precision, recall, and F1-score metrics commonly employed by the anomaly detection community. In particular, given negative and positive traces, the traces flagged as, respectively, normal and anomalous, the aforementioned metrics are calculated through the true positives (TP), true negatives (TN), false positives (FP) and false negatives (FN) classified by the technique:
\begin{flalign*}
\text{Accuracy} &= \frac{TP+TN}{TN+ TP + FP + FN}, &\\
\text{Precision} &= \frac{TP}{TP + FP}, &\\
\text{Recall} &= \frac{TP}{TP + FN}, &\\
\text{F1-score} &= 2 \cdot \frac{\text{Precision} \cdot \text{Recall}}{\text{Precision} + \text{Recall}}. &
\end{flalign*}
In addition, we also use the Area Under the Receiving Operating Curve (AUC). The ROC curve plots the \textit{True Positive Rate} (TPR) against the \textit{False Positive Rate} (FPR), defined respectively as:

\begin{flalign*}
\text{TPR} &= \frac{TP}{TP + FN},& \\
\text{FPR} &= \frac{FP}{FP + TN}.
\end{flalign*}

The AUC is then computed as the integral of the ROC curve:
\begin{flalign*}
\mathrm{AUC} &= \int_0^1 \text{TPR}(\text{FPR}) \, d(\text{FPR}),&\\
\end{flalign*}
and can be interpreted as the probability that the model assigns a higher anomaly score to a randomly chosen anomalous trace than to a randomly chosen normal one.

\begin{figure*}[!t]
\centering
\includegraphics[width=\textwidth]{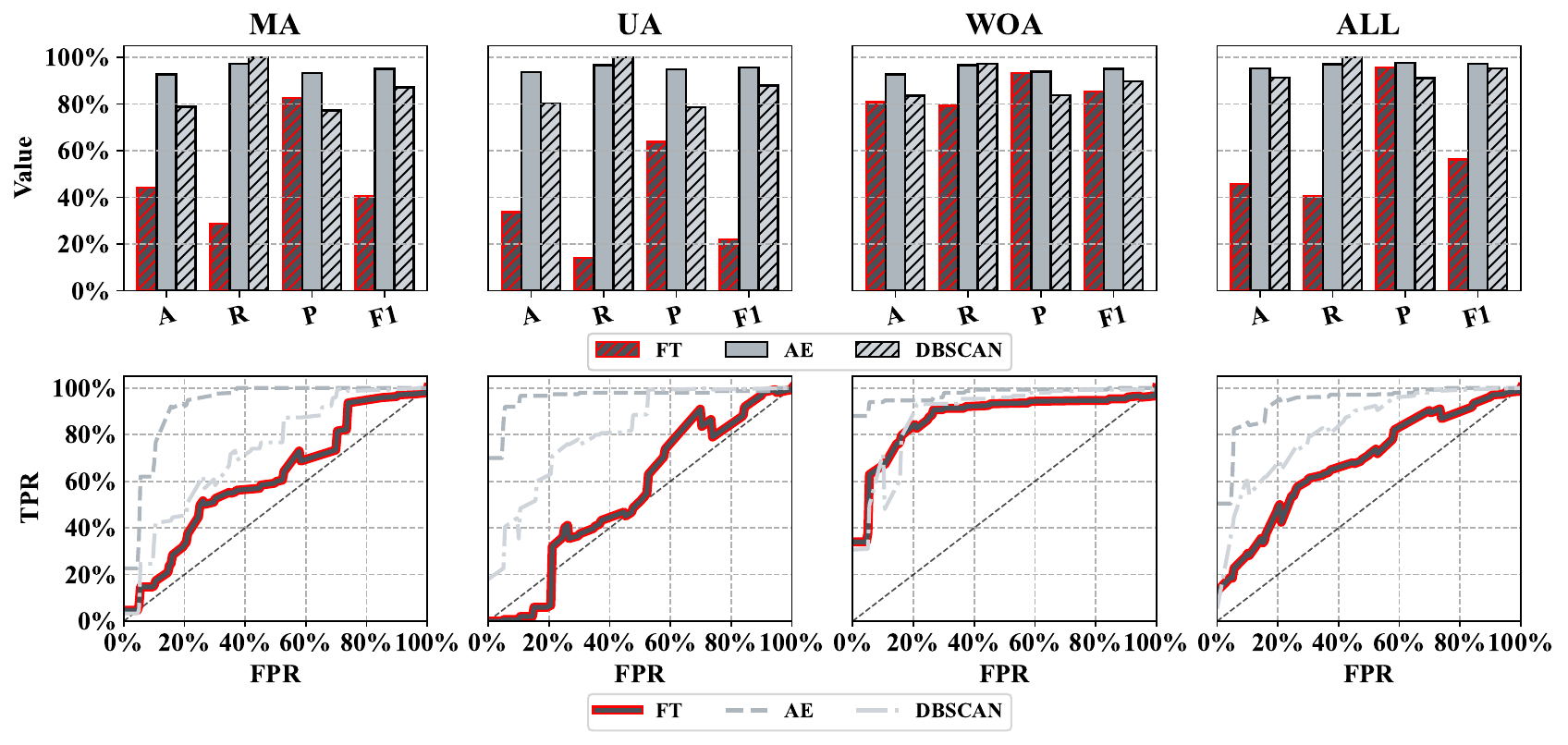}
\caption{Bar plots of the accuracy (A), recall (R), precision (P), and F1-score (F1), and ROC curves associated with each technique and anomaly type.}
\label{METRICS_PLOTS}
\end{figure*}
\paragraph{Data generation and experiment runs} The fault injection procedure is applied three times against the traces of an event log $L$ of 50 traces generated from the model in Figure \ref{SOM_PROCEDURE}. The first application of $f$ injects MA anomalies in the traces of $L$, resulting in $L_{MA}$. Similarly, the second and third applications result in $L_{WOA}$ and $L_{UA}$. In all three cases, we set $\lambda=3$, i.e., on average, each trace has three instances of the selected anomaly. Finally, we consider $L_{ALL}=L_{MA}\cup L_{WOA} \cup L_{UA}$.

\begin{figure}[!t]
\centering
\includegraphics[width=0.9\columnwidth]{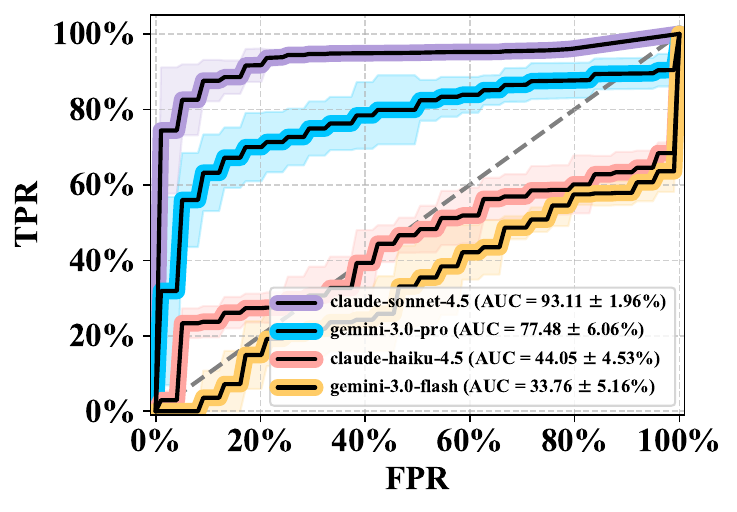}
\caption{ROC curves associated with the classification of normal and anomalous traces by the claude-sonnet-4.5's conformance checking–based monitor trained using the AE.}
\label{ROC_CURVE}
\end{figure}

\subsubsection{Results}
\paragraph{Intra-LLM validation}Table \ref{tab:AD_QUALITY} reports the results achieved for each conformance checking-based technique. The best-performing techniques are always either DBSCAN or AE, i.e., those that integrate unsupervised machine learning on top of conformance checking. They significantly outperform the baseline FT technique for the MA and UA anomalies, whereas the WOA anomaly is well discriminated by FT as well. This indicates that while fitness allows finding wrongly-ordered anomalies, it fails at identifying missing and unknown anomalies, for which an extension using other machine learning techniques is required. Considering the overall performance across all the anomalies, the AE is able to top the other two techniques, achieving up to 95.957\% F1-score and 93.669\% AUC. Figure \ref{METRICS_PLOTS} visually compares the metrics through bar plots and shows the corresponding ROC curves from which AUC is computed. The bar plots especially outline the superiority of the DBSCAN and AE techniques for all the anomaly types, especially for the accuracy, recall, and F1 metrics. The ROC curves at the bottom confirm this observation, and also show that the AE is the most stable across a wide variety of reconstruction error thresholds, i.e., its AUC is higher.

\paragraph{Inter-LLM validation} Figure \ref{ROC_CURVE} shows the ROC curves associated with the classification of normal and anomalous traces by the claude-sonnet-4.5's conformance checking–based monitor trained using the AE. The plot clearly shows that the event logs of claude-haiku-4.5 and gemini-3.0-flash lead to poor classification performance with AUC as low as 44.96\% and 35.22\%, respectively. On the other hand, while the classification of traces within the event log of gemini-3.0-pro is not as good as claude-sonnet-4.5, it still achieves a significantly higher AUC (79.17\%). These differences can be attributed to lower-quality instrumentation that deviates from the expected control flow, inducing Petri-net misalignments and increasing the false-positive rate. Hence, the preliminary analysis performed in the context of RQ1 is necessary prior to deploying the conformance checking–based software monitor.

\section{Discussion}
\label{sec:discussion}
\subsection{RQ1: Source-code instrumentation}
Obtaining quality software logs is pivotal to assessing the correctness of the computer-based system's execution. The main challenge in instrumenting code regards the adequate placement of log instructions and the connection of software traces to high-level models to ensure correct operation. We addressed this challenge by the systematic incorporation of LLMs in the engineering process of high-stakes systems. We have considered a set of LLMs that incorporate deep and shallow reasoning capabilities to evaluate their ability to understand both the source code and the process model of the prototype of SoM, the case-study ERTMS/ETCS scenario, and bridge the two by properly instrumenting the source code. The stimulation of these LLMs with an accurately engineered prompt aimed at preserving the code logic while instrumenting the code outlined satisfying results. In particular, the two LLMs with deeper reasoning (gemini-3.0-pro and claude-sonnet-4.5) achieved up to 82.849\% and 77.579\% coverage of the reference model's control flow.

\subsection{RQ2: Control-flow anomaly detection}
The ability to identify control-flow anomalies through the proposed conformance checking-based monitor has been evaluated with well-established anomaly detection metrics and different techniques trained on the diagnoses linked to event logs gathered by the best-performing LLM that resulted from the experimentation of RQ1. In particular, we have used the normal traces generated by the claude-sonnet-4.5 model to carry out our intra-LLM evaluation. The approach has proven successful, as different types of control-flow anomalies that can be linked to well-known software defects could be identified by the combination of alignment-based conformance checking and the AE with 95.957\% F1-score and 93.669\% AUC. This finding outlines the suitability of LLM-based source-code instrumentation for subsequent conformance checking-based control-flow anomaly detection. In addition, it is worth noting that using the AE does not invalidate the explainability of the proposed approach, as feature-based explanations can be provided through post-hoc explainability tools, such as SHAP, to identify the misaligned activities that caused the highest reconstruction errors \cite{vitale2025cfad}.

In addition, we have performed inter-LLM evaluation by comparing the ROC curves resulting from the classification of the event logs of the other LLMs, namely gemini-3.0-pro, claude-haiku-4.5, and gemini-3.0-flash. The results outlined that the source-code instrumentation quality is linked to the AUC metric, as gemini-3.0-pro achieved significantly higher AUC than the remaining two LLMs.

\subsection{Threats to validity}
Following common guidelines for empirical studies in software engineering \cite{wohlin2012experimentation}, we discuss potential threats to construct, internal, and external validity. Construct validity concerns whether the employed evaluation metrics adequately capture the intended phenomena. Internal validity addresses the extent to which observed results can be attributed to the studied factors rather than to uncontrolled variables. External validity concerns the extent to which the findings can be generalized beyond the specific context of the study.

\subsubsection{Construct validity}

We used a wide variety of metrics to evaluate both the instrumentation quality of LLMs and the ability of conformance checking to leverage LLM-based instrumentation for control-flow anomaly detection. While we believe that the quality of anomaly detection was comprehensively evaluated, our coverage metric focuses exclusively on the degree to which control-flow paths of the reference process model are exercised. Hence, it is worth mentioning that there are other, more traditional ways of assessing instrumentation quality, such as measuring \textit{code coverage} (e.g., statement, branch, or path coverage), \textit{instrumentation overhead} (i.e., the runtime and memory impact introduced by instrumentation), and \textit{trace completeness} (the proportion of relevant execution events successfully captured).

\subsubsection{Internal validity}  
Our experiments revealed that different LLMs, despite being instructed with the same prompt, produced varying results in terms of fitness, control-flow coverage, average trace length, and number of variants. While these differences can be partly attributed to the intrinsic architectural and training complexities of the selected LLMs, other factors may have contributed to the observed variability. For instance, aspects such as prompt formatting to adapt to the specific strengths and weaknesses of the given LLM, context length handling, or differences in the pretraining and fine-tuning data should be further explored. Finally, LLMs exhibit non-deterministic behavior that produces different outputs for the same prompt, unless constrained decoding settings are applied (e.g., low temperature, top-1 sampling, or greedy decoding). While we addressed the variation of source-code instrumentation through repeated and independent runs, a thorough sensitivity analysis on the parameters influencing LLM determinism could shed light on variability and the resulting quality of source-code instrumentation.

Regarding the anomaly detection experiments, although we employed both clustering-based and reconstruction-based techniques on top of conformance checking diagnoses, their performance might still be affected by other experimental factors. These include the specific fine-tuning procedures applied, the choice of hyperparameter configurations, and the stochastic nature of optimization processes.

\subsubsection{External validity} 
The transferability of our methodology to other domains is predicated on four prerequisites: 1) the availability of a design-time behavioral specification convertible into a formal model (e.g., the Petri net in Definition~\ref{def:petri_net}); 2) access to the system's source code for instrumentation; 3) a comprehensive test suite capable of establishing a reliable baseline of normal execution; and 4) an LLM with sufficient context window and reasoning capabilities to bridge the abstraction gap between the model and the implementation.

Regarding the generalizability of the ERTMS/ETCS results, broader validation is required. Future work should target real-world software implementations, specifically focusing on the detection of residual faults that are proven to escape standard verification strategies.
\section{Conclusions}
\label{sec:conclusions}
In this paper, we took on the problem of runtime monitoring of complex computer-based systems to detect control-flow anomalies due to unexpected changes in the system, the environment, or other ``unknown unknowns''. Given the limitations of runtime verification of declarative specifications in terms of strict consistency with prescriptive specifications and flexibility, we proposed a methodology to architect software monitors by combining LLMs with conformance checking to enable quality source-code instrumentation and fuzzy runtime control-flow analysis. We tested the methodology on a case-study scenario from ERTMS/ETCS, an industrial railway interoperability standard. The evaluation highlighted that the proposed methodology provides high-quality event logs through LLM-based source-code instrumentation and enables effective conformance checking-based control-flow anomaly detection.

Possibilities for industrial exploitation of the proposed approach include implementation through appropriate architectural frameworks for digital twins (see, e.g., \cite{flammini2021digital, 10049521}).
Future work will further investigate the efficacy of the methodology in real-world case studies, addressing the generalization of the proposal. In addition, a deeper investigation of the factors linked to LLMs that may influence the outcome of the instrumentation quality will be performed.

\section*{Acknowledgement}
The work of Francesco Vitale and Nicola Mazzocca was partly supported by the Spoke 9 “Digital Society \& Smart Cities” of ICSC - Centro Nazionale di Ricerca in High Performance-Computing, Big Data and Quantum Computing, funded by the European Union - NextGenerationEU~\seqsplit{(PNRR-HPC, CUP: E63C22000980007)}.

The work of Francesco Flammini was partly supported by the Swiss State Secretariat for Education, Research and Innovation (SERI) under contracts no. 23.00321 (Academics4Rail) and 24.00528 (PhDs EU-Rail); those projects have been selected within the European Union’s Horizon Europe research and innovation programme under grant agreements no. 101121842 and 101175856, respectively.

Views and opinions expressed are however those of the authors only and do not necessarily reflect those of the funding agencies, which cannot be held responsible for them.

\bibliographystyle{elsarticle-harv} 
\bibliography{bibliography.bib}

\end{document}